\RequirePackage{fix-cm}
\documentclass{svjour3}                     
\DeclareUnicodeCharacter{0301}{\'{e}}
\smartqed  
\usepackage{graphicx}
\usepackage{microtype}
\usepackage{enumitem}
\usepackage[utf8]{inputenc}
\usepackage[british]{babel}
\usepackage{tabularx}
\usepackage{multirow}
\usepackage{makecell}
\usepackage{longtable}
\usepackage[utf8]{inputenc}
\usepackage{hyperref}
\usepackage{float}

\usepackage{csquotes}
\usepackage[natbib=true,style=authoryear-comp,hyperref,backend=biber,isbn=true,doi=true,url=true]{biblatex}

\usepackage{lipsum}
\usepackage{booktabs,amsfonts,dcolumn}
\newcolumntype{d}[1]{D..{#1}}

\usepackage[table,xcdraw]{xcolor}
\usepackage[normalem]{ulem}
\useunder{\uline}{\ul}{}

\bibliography{0-TrustSECO_SLR}

\begin{document}

\title{A Systematic Literature Review on Trust in the Software Ecosystem}


\author{Fang Hou         \and
        Slinger Jansen 
}


\institute{Fang Hou \at
              Department of Information and Computer Science, Utrecht University, the Netherlands \\
              \email{h.fang@uu.nl}           
           \and
           Slinger Jansen \at
           Department of Information and Computer Science, Utrecht University, the Netherlands\\
           School of Engineering Science, Lappeenranta University of Technology, Finland
              \email{slinger.jansen@uu.nl}
}

\date{Received: date / Accepted: date}

\maketitle

\begin{abstract} 
\hfill \break
{\noindent \textbf{Background:}  
The worldwide software ecosystem is a trust-rich part of the world. Throughout the software life cycle, software engineers, software end-users, and other stakeholders collaboratively place their trust in major hubs in the ecosystem, such as package managers, repository services, and software components. However, as our reliance on software grows, this trust is frequently violated by bad actors and crippling vulnerabilities in the software supply chain, endangering our livelihoods and the resilience of our society.

\noindent\textbf{Purpose:} The purpose of this study is to define trust in the worldwide software ecosystem, that is, to determine what signifies a trustworthy system or actor. By improving our understanding of trust, we can create novel mechanisms for protecting the software ecosystem from vulnerabilities and attacks.

\noindent \textbf{Methods:} We conduct a systematic literature review on the concept of trust in the worldwide software ecosystem. We acknowledge that trust is something between two actors in the software ecosystem, and we examine what role trust plays in the relationships between end-users and (1) software products, (2) package managers, (3) software producing organizations, and (4) software engineers.

\noindent \textbf{Results:} Two major findings emerged from the systematic literature review. To begin, we provide a definition of trust in the software ecosystem, including a theoretical framework that decomposes and signifies a theoretical understanding of trust. Second, we provide a list of trust factors that can be used to assemble an overview of software trust.

\noindent \textbf{Conclusion:} Trust is critical in the communication between actors in the worldwide software ecosystem, particularly regarding software selection and evaluation, and can affect the outcome either positively or negatively. According to the frequency of the trust factors in the literature, the most significant determinants of trust are software quality, code and structure, and the reputation of software producing organizations. With this comprehensive overview of trust, a new foundation is laid for software engineering researchers to understand and use trust to create a reliable software ecosystem.}

\keywords{Software ecosystem \and Software trust \and Software security \and Software package evaluation}

\noindent\textbf{Statements and Declarations}
 
\noindent The authors have no relevant financial or non-financial interests to disclose.

\noindent No funding was received to assist with the preparation of this manuscript.

\noindent Author Slinger Jansen is an Editorial Board Member of Empirical Software Engineering.

\end{abstract}

\section{Introduction}\label{sec:intro}

When software engineers are building software, for example, in-house software, they are constantly being confronted with a choice: build it themselves or integrate a component from another software producing organization. If the choice is made to integrate a component, the second choice is a selection problem: there are typically many options to select from and it depends on the software engineer's perspective which factors heaviest in the selection process. Factors such as feature completeness, software component quality, and technology can all determine whether the software engineer selects a component, while researchers have called for more structured approaches to software selection~(\cite{farshidi2021decision}), the selection process is typically completely dependent on the context of the software engineer. 

The software ecosystem is a trust-rich part of the world. End-users expect a certain trustworthiness when downloading apps from their favorite application store. For instance, when the appstore provides checksums, a technology to determine the authenticity of the received data, for each deliverable, the end-user can ensure that the data she receives is the same data that came from the software engineer. Nevertheless, it is more often than not the case that trust is put into Software Ecosystem (SECO) hubs such as package managers and appstores, without there actually being a good basis for this trust. Trust can be considered as founded and unfounded. There are also more soft ways to create founded trust, e.g., ensuring that the software engineer has been a productive member of the SECO for a long time, or ensuring that the software developed achieves established quality levels. Both soft and hard ways do not provide watertight guarantees, as the external environment can also have an impact on trust, e.g., new attacks from the external world. According to~\citet{Sonatype2021OSS}, the number of supply chain attacks is dramatically increasing, posing significant security risks to both software producing organizations and end-users. For example, bad actors shift their attacks ``upstream'' by inserting malicious code to the source code repositories to acquire a critical advantage of time, allowing malware to spread throughout the supply chain. 

Since we have already mentioned trustworthiness and trust, it is necessary to explain their differences, as we found that trust and trustworthiness are used interchangeably, and often trustworthiness is mistakenly interpreted as trust~(\cite{wright2010trust}). \textbf{Typically, trust is from the trustor's perspective, it is a belief that the trusted person will actually do what is expected of her}~(\cite{jadhav2011framework,guo2014ratings,mcknight2011trust,bauer2019conceptualizing}), while \textbf{the concept of trustworthiness can be viewed as a probability, i.e., the probability that a trustee will act as expected by the trustor}~(\cite{bauer2019conceptualizing}). Hence, a trustee's trustworthiness and a trustor's trust may differ. In addition, it is found that trustworthiness can be seen as a characteristic of trust~(\cite{becerra2008trustworthiness}), which means trust cannot exist without a high level of trustworthiness~(\cite{heyns2015dimensionality}). 

Extensive research has shown that trust is crucial in software selection and evaluation~(\cite{bugiel2011scalable}). 
Typically, end-users trust in software is built upon the two following aspects: one is the  product should perform its desired functions and meet their specific needs; the other is the software producers should provide reliable services. Software trust differs from software ecosystem trust in that it concentrates on the software products and their producers. More information on software trust and software ecosystem trust can be found in Section \ref{STdef}.

Currently, standards and models have been proposed to assess software trust, for instance, the Trusted Computer System Evaluation Criteria~(\cite{10.5555/1593439}) and the Reliability Growth Model~(\cite{littlewood1994learning}). However, they are not widely used. The primary reasons are : 
\begin{itemize}
\item In light of the differences in the needs of end-users for various software products, these models lack comprehensive and uniform criteria for measuring the trust in all software products~(\cite{yan2008comprehensive}); 
\item Most of the factors are abstract and lack clarity or guides that can be used to quantify trust factors~(\cite{li2021exploring}); 
\item There is a lack of breadth of perspective among the criteria for judging the trustworthiness of software that practitioners would expect. For example, we rarely receive ratings for various package versions and a few comparable software producer ratings. Yet, this is precisely what software practitioners are concerned about when assessing software quality~(\cite{goode2015rethinking}). 
\end{itemize}

In this context, we conduct a systematic review of the literature (SLR) to better understand software trust in the software ecosystem and summarize the key impact factors based on the frequency with which they were mentioned in the studies. The SLR presented here is part of a larger study of software trust that breaks a lance by giving trust a proper role in this selection process. By analyzing the perception and trust of various stakeholders, we provide a comprehensive overview of the relevant trust concepts. The purpose is to lay the groundwork for research on the design of trust assessment mechanisms that will place trust at the center of the selection process for software engineers.

\begin{figure*}[!htbp]
    \centering
    \includegraphics[width=1.7\textwidth,angle=90]{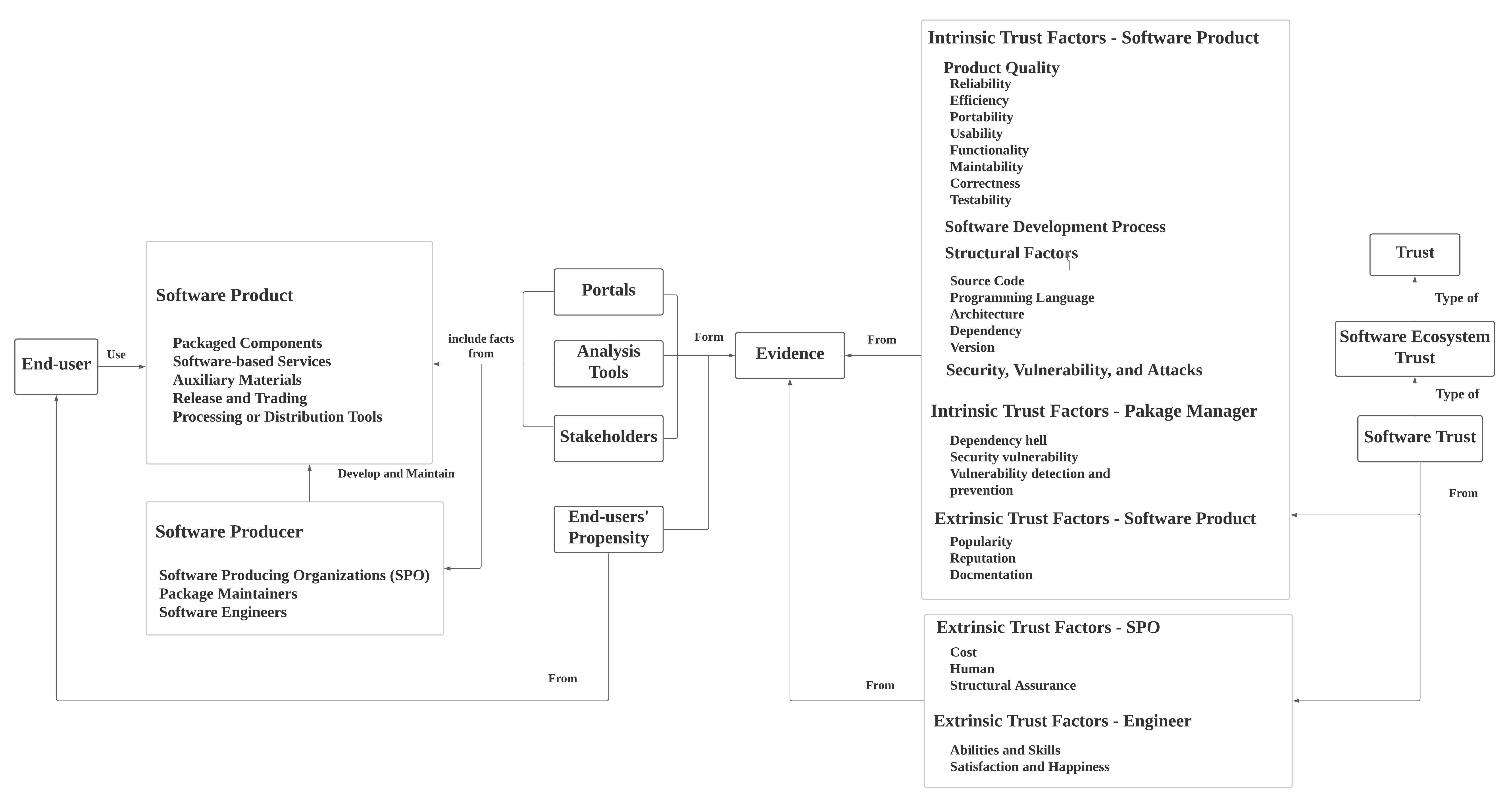}
    \caption{\footnotesize\textit{The meta-model can be used as a reading guide for this article, as it mentions all the relationships between the main entities in our research. For instance, the relationships between trust types, trust sources, major trust impact factors, hubs, and stakeholders in the software ecosystem.}}
    \label{fig:meta-model}
\vspace{-70pt}
\end{figure*}

To help understand this study, we develop a model in Figure \ref{fig:meta-model} to depict the relationship between significant trust impact factors and actors in the software ecosystem. It can serve as a brief overview of the results of this study, a detailed discussion of which is provided in Section \ref{sec:Results}.

This research begins with the following outline:
\begin{description}

\item[$\bullet$] Section \ref{sec:background} introduces the actors and hubs involved in software selection and evaluation in the software ecosystem, as well as the workflow of the software packages management. The purpose is to identify relevant trust impacted elements in the software ecosystem;

\item[$\bullet$] Section \ref{sec:ResearchMethod} presents the research method we used in this literature review, which works as an acquisition process to collect the software trust concept and trust factors considered by the software end-users during the software selection; 

\item[$\bullet$] Section \ref{sec:Results} provides a definition of software trust as well as a discussion on the intrinsic and extrinsic trust factors in terms of software components, package administrators, software production organization, and engineers, respectively. The results show that quality is the decisive factor for software trust, followed by software producers and code \& structure. Additionally, trust assessment models typically focus on security, such as data access or user entitlement, rather than vulnerabilities and attacks; 

\item[$\bullet$] Section \ref{sec:dission} highlights the effect of trust factors on the software selection, validity, consideration, and challenges of this work, as well as our future work. It is difficult to develop a consistent set of criteria for assessing software trust; therefore, our view is that trust assessment should be synthetic and drawn on from different perspectives and channels. When collecting and sharing information, we need to exercise caution to protect both the privacy of individuals and the objectivity of the data. Additionally, collecting data on the trust of closed software may be challenging.
\end{description}

Finally, in Section \ref{conclusion}, we conclude that current research on software trust is limited to software quality, ignoring the impact of other hubs or actors in SECO, e.g., software packages, package managers, or software producers the future research should expand on this topic.

\section{Background of Software Ecosystem}
\label{sec:background}

To assist in understanding what aspects of SECO can affect software trust, we present here an overview of the package management workflow, as well as the concepts used in the study. We sketch the primary entities and stakeholders in the software ecosystem in Figure \ref{fig:PM workflow}, along with their simplified relationships.

\subsection{Actors in the Software Ecosystem}
\noindent \textbf{Software Product} - The definition of the software product is based on Xu and Brinkkemper's concept of a software product. It is defined as \textit{``a packaged configuration of software components or a software-based service, with auxiliary materials released for and traded in a specific market''}~(\cite{xu2007concepts}). The definition emphasizes four elements: ``\textit{packaged components}'' refer to software, packages, components, libraries, or codes; ``\textit{software-based services}'' refer to the independent, small piece of functionality accessed through internet, ``\textit{auxiliary materials}'' refer to documentation, user manual or other materials which need to be implemented with the software or delivery to the end-users , and ``\textit{release and trading}'' refer the activities that release the software to the market and related commercial value. To this, we add another independent element -``\textit{distribution tools}'', emphasize channels that facilitate distribution downstream of the software supply chain, meaning the systems or tools that assist in the processing or distribution of the packaged components that comprise a software product, e.g., package managers. 

\noindent \textbf{Component} - Based on Software Engineering Body Of Knowledge (SWEBOK), \textit{``a software component is an independent unit, having well-defined interfaces and dependencies that can be composed and deployed independently''}~(\cite{Bourque2014swebok}). The significant characteristic of a component is that it can be reused, can interact with and interact with other objects, and can be combined with other components to form a system or application. For instance, a menu class or a button class.

\noindent \textbf{Library} - A library is a collection of prewritten functions or data structures that are organized to perform the same technically essential functions, such as functions that handle compatibility issues~(\cite{bauer2012structured}). Sometimes a library is referred to as a package in programming language directives, such as npm, RubyGems, and Maven~(\cite{zerouali2018empirical}).

\noindent \textbf{Package} - A package is simply a collection that contains software, libraries and metadata. The metadata includes, for example, the name of the software, a description of its purpose, version number, providers, checksum, and a list of dependencies for the software or library version. Packages are in general versioned, which frequently adhere to de facto conventions, such as semantic versioning~(\cite{hanus2018semantic}). A distribution of packages can be considered as being a software ecosystem, with a collection of interdependent software projects that are developed and maintained within the same environment~(\cite{decan2019package}). 

\noindent \textbf{Package Manager} - A package manager provides a privileged, central mechanism for managing the installation and upgrade of packages on a computer's operating system automatically~(\cite{Cappos2008alook}). Sometimes it is referred to as Software Manager or Application Manager, depending on the context. The preferred term is Package Manager in programming languages since the installed software is often just a set of libraries rather than a directly executable application. Package managers have a major contribution in that they retrieve the specific versions of the libraries that are required to build client applications and the libraries they depend on, and install the required libraries based on their dependencies~(\cite{hejderup2018software}).

\noindent \textbf{Package Repository} - Typically, a package repository is just a web server that hosts packages and their associated metadata~(\cite{cappos2008package}). Package managers rely on package repositories to install packages and resolve any dependency requirements.

\noindent \textbf{End-user} - A software end-user is an individual or organization that adopts or intends to adopt a software product into their applications. Usually, the majority of end-users are software engineers, who are concerned with the functionality of a package and its integration into their existing projects~(\cite{chinthanet2021makes}). Alternatively, they could be individuals who lack development experience and rely on a software component to fulfill a specific requirement. The distinction between these two types of end-users is not emphasized in our study. However, in our literature review, we found that most of the interview or survey participants are experienced software engineers.

\noindent \textbf{Software Producer} - The term software producer refers to both software producing organizations (SPOs) and software engineers who are involved in its development. An SPO is an organization that builds and maintains software with a view to getting the software created, adopted as widely as possible~(\cite{jansen2012shades}). As software engineers, whether they are employed by an SPO or as independent contractors, they contribute to the development and maintenance of software products.

Additionally, package maintainers are a subset of software producers. Package maintainers are responsible for the development and maintenance of software packages and their frameworks. They develop and maintain source code, package builders, and installers, including bug fixes and patch releases. Additionally, they configure package builds to ensure that packages can be sourced from the distribution by extracting the source code from the collection of binaries in the distribution~(\cite{Duan2020Measuring}). 

\subsection{Workflow of Package Management}
\begin{figure*}[!htbp]
    \centering
    \includegraphics[trim=20 170 20 150,clip,width=0.9\textwidth]{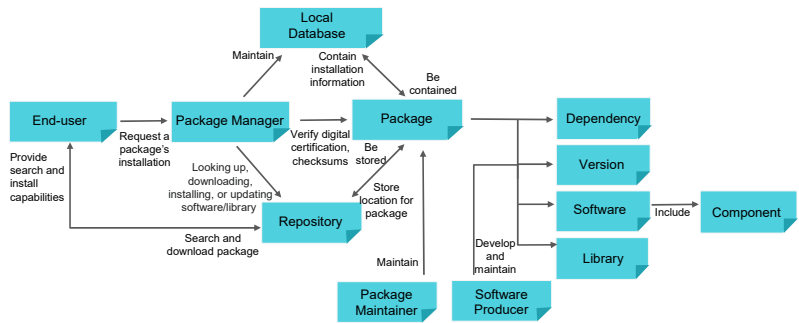}
    \caption{\footnotesize\textit{The model shows the package management process, and it includes relevant stakeholders, software end-users and package maintainer, essential entities, package manager, package, local database, and repository.}}
    \label{fig:PM workflow}
\end{figure*}

In Figure \ref{fig:PM workflow} we describe the package management process.  When an end-user uses a package manager to download a library, first, the package manager searches the relevant configuration file maintained by itself for files that point to the location of the repository, e.g., Apple Store or SourceForge, to retrieve the package. Subsequently, the package manager downloads the relevant packages from the repository. When the end-user confirms the installation of the selected package, the package manager installs the relevant packages according to the package dependency order. After a successful package installation, the package installation information is stored in the metadata of the local package database and managed by the package manager to maintain software dependency and version information. End-users can modify the configuration file to retrieve packages from other repositories. Alternatively, they can search and download packages directly from repositories, such as Github.

A trustworthy program or application cannot depend on untrustworthy components or services~(\cite{Bourque2014swebok}). To this end, in addition to the trustworthiness of the software product itself, we should also be concerned about the reliability of the reusable components, such as components and libraries, as well as the the trustworthiness of the tools that distribute or store them, such as packages and package managers, and repositories that host these reusable components. These elements are taken into account in the study of software trust.

\section{Research Method}
\label{sec:ResearchMethod}

Systematic Literature Review (SLR) is a term used to describe the process of collecting, reading, analysing, refining, and organizing of data in the existing literature to provide a comprehensive introduction, elaboration, and evaluation on a specific research topic or phenomenon of interest~(\cite{keele2007guidelines}). We performed an SLR following the guidelines and steps of~(\citet{kitchenham2004procedures}) to gain the knowledge in the existing literature on the concepts and influences of trust within the software ecosystem domain to understand and reveal the current relationships between software end-users and software producers. It followed the outline of the following six activities in this review:

\begin{itemize}
\item Defining the research question; 
\item Searching for relevant studies; 
\item Applying inclusion and exclusion criteria; 
\item Assessing study quality;
\item Extracting and analyzing data;
\item Coding scheme.
\end{itemize}

\subsection{Research Questions}
We construct the following research questions (RQ) to get an overview of software trust in worldwide SECO:
\\
\newcommand{\RQOne}{How is the concept of software trust defined in literature?}
\newcommand{\RQTwo}{What trust factors do end-user organizations consider when selecting software products?}
\newcommand{\RQTwoDotOne}{What trust factors do end-user organizations consider when selecting software products and versions?}
\newcommand{\RQTwoDotTwo}{What trust factors do end-user organizations consider when selecting software package managers?}
\newcommand{\RQTwoDotThree}{What trust factors do end-user organizations consider when selecting software producing organizations?}
\newcommand{\RQTwoDotFour}{What trust factors do end-user organizations consider when selecting software engineers?}

\noindent \textbf{RQ1: \RQOne}

\begin{itemize}
\item[]
We define the concept of software trust as the first object of this study. This is achieved by reviewing and comparing the definitions of software trust in the literature and examining the sources, types, and attributions of software trust. In addition, we extend the definition of software trust based on the relationships between software end-users and producers in the software ecosystem. 
\end{itemize}

\noindent \textbf{RQ2: \RQTwo}

\begin{itemize}
\item[]
RQ2 takes into account end-user organizations' trust factors when making software product selection decisions, which can be regarded in the following four sub-research questions. Their impact on software trust may be positive or negative, in varying degrees.
\end{itemize}

\noindent \textbf{SubRQ2.1: \RQTwoDotOne}

\begin{itemize}
\item[]
RQ2.1 addresses the factors in software and software packages that affect trust. These factors are comprehensive in that part of them can originate from the software and software packages themselves, primarily the quality of the source code (e.g., integrity, re-usability), known defects and vulnerabilities, packages' versions and dependencies, and the development process that affects the quality and security of these source codes; another part of them can come from the outside world, such as the reputation, popularity, and structural assurance of the software.
\end{itemize}

\noindent \textbf{SubRQ2.2: \RQTwoDotTwo}
\begin{itemize}
\item[]
RQ2.2 is concerned with package managers. Since the major role of package managers is to manage dependencies in order to ensure that packages are installed with all dependencies they require, the ability to manage package dependencies becomes a critical factor affecting the trust in package managers. Additionally, the trust in package managers is also affected by security risks, such as consistently detected attacks on metadata, dependencies, and repositories, as well as known design flaws, such as the lack of effective protection mechanisms to detect and undo suspicious package installations.
\end{itemize}

\noindent \textbf{SubRQ2.3: \RQTwoDotThree}
\begin{itemize}
\item[]
RQ2.3 focuses on how software producers, organizations/ communities build trust in software and software packages. It focuses primarily on their management and strategy, such as follow-up of their products, reputation, and popularity.
\end{itemize}

\noindent \textbf{SubRQ2.4: \RQTwoDotFour}
\begin{itemize}
\item[]
RQ2.4 is concerned with the professionalism of software engineers, especially their knowledge, experience, and skills, as these factors all contribute to their ability to develop high-quality software, effectively resolve unexpected problems, and provide follow-up services. Engineers’  satisfaction and happiness are also beneficial for developing quality software. 
\end{itemize}

\subsection{Search Strategy}

Literature review articles are identified by using search strings from scientific libraries. First, we define the object of study, including software and related concepts, such as software packages, components, and dependencies. Then, in conjunction with software supply chain processes, we include software management, provenance, and engineers, considering that software trust may be influenced by project management and other human factors. Finally, after examining a few definitions of trust, we include concepts related or similar to trust based on the understanding of the nature of trust, such as credibility, reputation, and uncertainty. This resulted in the following block.
\\

\fbox{%
\parbox{0.9\textwidth}{%
\begin{flushleft}
software AND (package OR component OR dependency OR management OR provenance OR engineer) AND (trust OR credibility OR reputation OR uncertainty)
\end{flushleft}
}
}\leavevmode\newline

\begin{figure*}[!htbp]
    \centering
    \includegraphics[width=\textwidth]{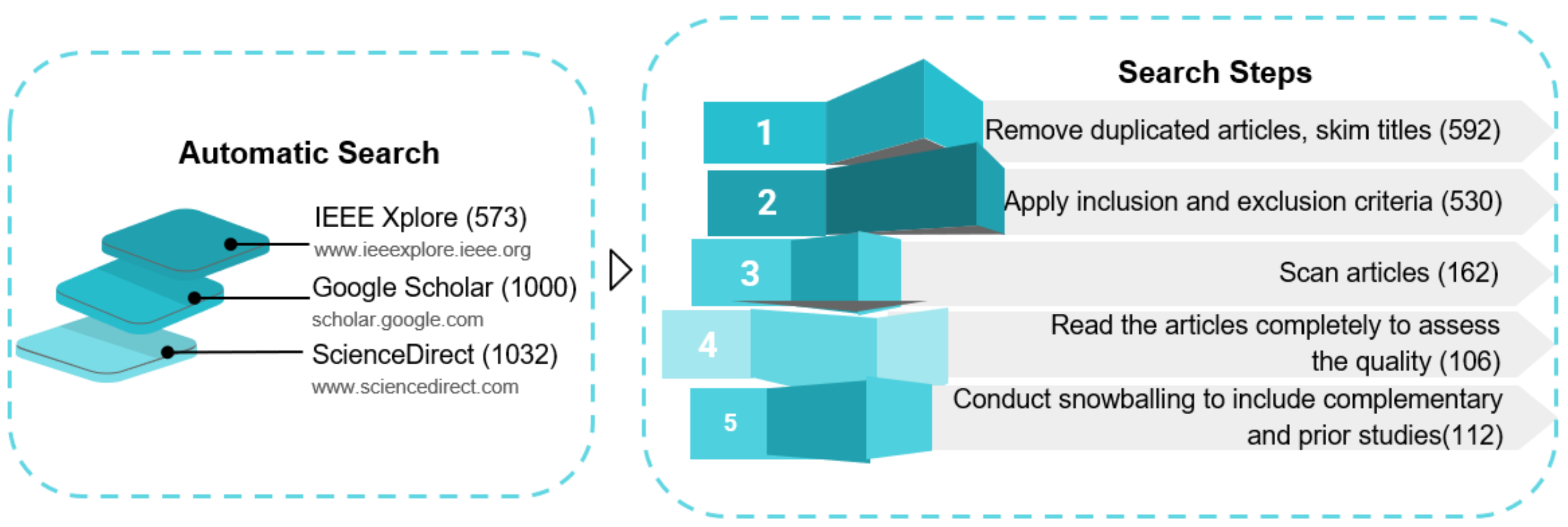}
    \caption{\footnotesize\textit{This figure indicates the stages of the search process and the number of publications at each stage.}}
    \label{fig:search process}
\end{figure*}

Software trust has been studied within software engineering, and related research can be found in any software engineering-related digital library. To ensure that no studies were missed, we adopted both manual search, automatic search, and snowballing technique. We began by manually searching for primary and secondary studies from the Journal of Systems and Software and Information and Software Technology. However, it was found that the articles we searched manually were included in the automatic search results, and they also incorporated a wider range of articles. Thus, we decided to use automatic search as the main search method, kept automatic search results from IEEExplore and ScienceDirect as the academic sources, as well as Google Scholar as a secondary source. Then, the first author obtained the initial list of articles from the search engine and eliminated duplicates. 592 articles were chosen based on their titles, excluding articles that clearly do not belong to software engineering, for example, we encountered titles containing community environmental health, fish, and patients. In the following process, five members of our research team applied inclusion and exclusion criteria to each article, and if all the criteria were met, the article was then included. After that, the five researchers scanned the remaining articles to exclude more articles, then read the articles completely, and performed the quality assessment. Upon reading the articles in their entirety, we used a snowballing technique to include another seven articles related to our topic, such as supplements and prior research, ensuring that we did not miss any extensive versions of the articles. The steps of the search process are depicted in Figure \ref{fig:search process} along with the number of primary and secondary studies conducted at each step. Note that there is considerable overlap between articles in different digital libraries, resulting in the inclusion of numerous irrelevant studies. 

\subsection{Inclusion and Exclusion Criteria}

The inclusion and exclusion criteria ensure that relevant studies are included, and studies that do not answer research questions are excluded. If the inclusion criteria are too broad, poor-quality studies may be included, lowering the overall quality of the study's results. If the inclusion criteria are too stringent, the resulting studies are likely to be small and therefore not generalizable(\cite{meline2006selecting}). 

\noindent The given inclusion criteria adopted in this research are:
\begin{itemize}
 \item Studies published since 1990s; 
 \item Studies that addressed at least one research question; 
 \item When studies reported the same research, only the most recent one was included;   
 \item Studies that appropriately addressed software trust in software ecosystems.
\end{itemize}

\noindent The given exclusion criteria selected in this research are:
\begin{itemize}
 \item Studies that published in a language other than English;
 \item Studies that incomplete or only provided literature in the form of abstracts or presentation slides;
 \item Studies that concentrated on the development of a single tool for a particular domain;
 \item Studies that lacked an adequate description of their objectives or context;
 \item Studies that were not specifically related to software engineering or software ecosystem.

\end{itemize}

We include books, articles, and a variety of grey literature, e.g., working papers or dissertations/theses, on software trust, software selection, and software evaluation. These search results were published after the 1990s, as the Internet revolutionized the relationship between people, as well as the relationship between people and software since then. Meanwhile, to avoid results from other disciplines interfering with this study, we excluded search results that did not pertain to software engineering or software ecosystems, such as  software packages that are explicitly applied to medical or mathematical sciences. Additionally, we excluded search results that were either incomplete or lacked detailed descriptions. 

\subsection{Quality Assessment}
The quality assessment guides the interpretation of the findings~(\cite{keele2007guidelines}). In this study, five researchers were involved in the quality assessment by adopting the double data extraction method~\citep{buscemi2006single}, which means the quality of each article was assessed by two researchers. We organized discussions to reach a consensus on quality assessment and read three articles to cross-check and compare the results to use this mechanism to eliminate any doubts about the selection process. Our quality assessment focuses on the following questions:

QA1. Does the study address at least one research question?

QA2. Is the study based on research or expert opinion?

QA3. Is there a clear statement of the purpose of the study?

QA4. Was the data collection rigorous?

QA5. Are the findings of the study clear?

\noindent Each member of the research team was then randomly assigned a subset of these 592 articles. The first author then conducted a quality assessment of all researchers' derived results, discovering that 28 articles received varying quality assessment scores. These 28 articles included six Google Patents, four articles about specific software systems, such as mobile applications or enterprise resource planning, and 19 articles about topics other than software trust, such as hardware and internet trust, or how to estimate maintenance. Following a discussion of these distinctions, we arrived at 112.

\subsection{Data Extraction and Synthesis}
\label{datasets}

The research methods were classified according to the categories listed in Table \ref{researchmethods}. The majority of the search results are empirical research. Theories, models, or frameworks that analyze or evaluate software trust or specific aspects of software products contributed most of the trust factors to our study. The remaining factors came from theoretical studies and exploratory research, e.g., literature review. The results can be seen in Table \ref{datacollec}.

Two data sets have been created to compile the impact factors found in the literature. 

\textbf{The qualitative data set} - The analysis of the literature or its findings can be used to deduce the factors affecting software trust. The majority of significant factors are listed in tables or explicit categories, while others were analyzed and extracted from the literature's context, findings, and conclusions. 

\textbf{The quantitative data set} - The impact factors for software trust are determined from the models, frameworks, formulas, or test results shown in the search results. The quantitative data set was composed of works that employed experimental and statistical data collection methods. Typically, the impact factors are presented graphically in tables, or as formulas, or they are classified within the discussion and context of such literature. 

\begin{table}[!htbp]
\centering
\begin{minipage}[t]{0.48\linewidth}\centering
\caption{Summary of types of the search results}\label{researchmethods}
\begin{tabular}{ c r }
\toprule
Research types & Total \\
\midrule
Empirical Research & 83 \\ 
Theoretical Research & 18 \\ 
Exploratory Research & 10 \\
\bottomrule
\end{tabular}
\end{minipage}\hfill%
\begin{minipage}[t]{0.48\linewidth}\centering
\caption{Methods of data collection used in the empirical research}\label{datacollec}
\begin{tabular}{ c r }
\toprule
Data collection methods    &  Total \\
\midrule
Experiment & 42 \\ 
Case Study & 19 \\ 
Survey/Interview & 18 \\ 
Statistic & 4 \\ 
User study & 1 \\ 
\bottomrule
\end{tabular}
\end{minipage}
\end{table}

Data synthesis was performed using a frequency analysis technique to aggregate the extracted data. All included articles were analyzed and extracted. It should be noted that because some models or criteria have been proposed based on the results of previous research, several trust factors are duplicated. In the data synthesis, we still counted and accumulated these trust factors. The reason is that we believe the trust factors that are repeatedly referred to must be significant. Section \ref{sec:Results} presents the synthesis results of the data extracted from the selected studies and answers the research questions.

\subsection{Coding Scheme}
\label{codingscheme}

To analyse what trust factors end-user organizations consider when selecting software products, we classified the impact factors and tried to find the relationships between them. We adopted inductive coding method to analyze the data from each study. Inductive coding is a type of data analysis in which the researcher reads and interprets raw textual data to develop concepts, themes, or a process model based on the data~\citep{chandra2019inductive}. In addition, considering our topic, software trust, used combining Cue Utilization Theory (CUT) to categorize trust factors~\citep{midha2012factors}. As CUT is well-known as a framework for understanding and analyzing the different factors that influence the subject matter as well as evaluating products. We classified the cues we selected from the search results using CUT and divided the impact factors into intrinsic and extrinsic trust factors for the previously proposed research questions. Intrinsic trust factors have been used to represent a product's physical attributes, which refer to source code or architecture-related attributes or factors, such as technical factors or vulnerabilities; and extrinsic trust factors have been used to represent external attributes, such as the product's reputation, cost, licenses, or capability of the software producer. In addition, we focus on each actor in the SECO and their relationships, for example, the reputation or popularity of software producers, which are also categorized in extrinsic trust factors.

After conducting a systematic review of all impact factors, we assigned them to our CUT categories as well as to the research questions based on the themes of each study. We then labeled the codes using the words from both datasets, as well as their CUT and semantic counterparts and then adapted and refined them. Additionally, we determined the relative importance of each category by counting the frequency of the words. If a theme appears frequently, it indicates that it is significant in the research on software trust and may have a significant impact on software trust. In Figure \ref{fig:factors} we show the major trust impact factor according to the word frequency. The coding scheme of trust impact factors is available at
\href{https://data.mendeley.com/api/datasets/xn4jk93g4t/draft/files/b36ddc93-0f42-4227-b966-4c07d2a7e429?a=9c51b45f-c943-4a22-814f-5baf59fb2896}{this link}. 

\subsection{Biases in the SLR Process}

Bias is one of the biggest challenges faced by all researchers. For example, researchers may be unaware of previous research on a subject or maybe aware but unable to access the findings, or the report may be missing critical data points. Thus, when data are missing, not only does the size of the collected sample decrease, but the sample's validity gets smaller as well~(\cite{cooper2008research}). In this study, we faced two sampling biases: \textbf{Publication bias} refers the problem that positive results tend to be published more frequently than negative results, if negative results are not published widely enough, false conclusions could be attributed to them as true~(\cite{keele2007guidelines}), and \textbf{Retrieval bias}, Typically, retrieval bias refers to the risk that the sample used in the synthesis does not appropriately represent the literature available~(\cite{Dur2017bias}). In order to mitigate them we did the following:

\begin{itemize}

\item Conduct both automatic search and snowballing technique. The automated search we conducted includes a broader search of articles about software trust, which is supplemented by snowballing techniques to make the search more comprehensive in content and scope.

\item Obtain articles from published and unpublished articles. In this study, along with articles from peer-reviewed journals, we included grey literature, such as pre-prints and post-prints of articles, theses, and dissertations.

\end{itemize}


\section{Results}
\label{sec:Results}

\begin{figure*}[!htbp]
    \centering
    \includegraphics[width=\textwidth]{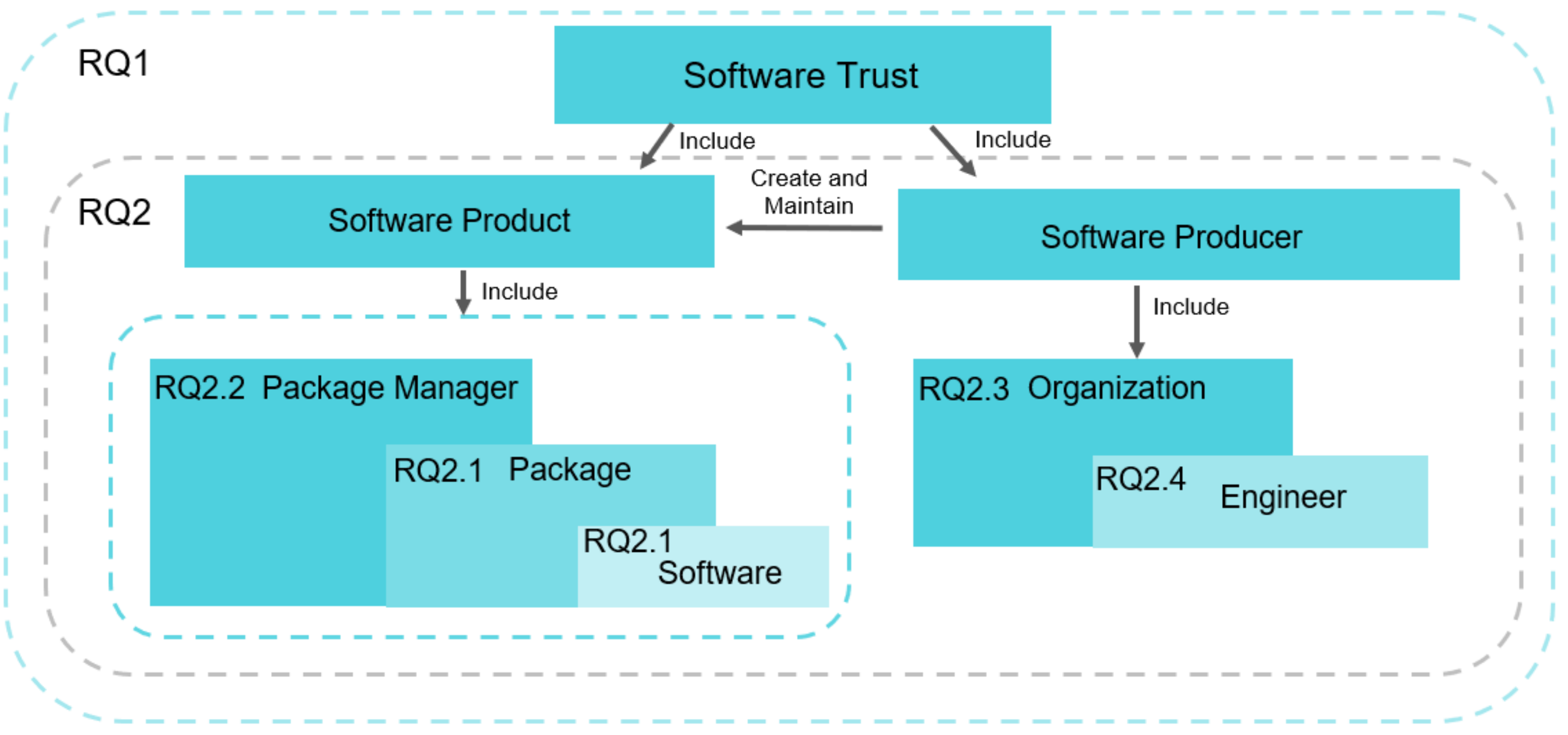}
    \caption{\footnotesize\textit{This model shows the structure of this study, including the relationship between the research questions.}}
    \label{fig:rq}
\end{figure*}

This section reports our findings of software trust. Figure \ref{fig:rq} illustrates the structure and relationships of our research questions, thereby contributing to our overall understanding of our study. It also shows the order we answer the RQs. First, we provide the definition and attributes of software trust (RQ1) to understand its underlying constructs (RQ2), which are software products and software producers. Following this, we examine the software products and their producers to discover the factors that may influence software trust in the SECO during the selection process. In terms of software products and software producers, we analyze the trust factors affecting software and software packages (RQ2.1) and package administrators (RQ2.2), as well as software production organizations (RQ2.3) and engineers (RQ2.4). Numbers of search results for each research question are given in Table \ref{papers by rq}. 

\begin{table}[!htbp]
 \begin{minipage}{\textwidth}   
\centering
\caption{Overview of selected studies per research question. It is possible for a publication to address more than one research question.}
\label{papers by rq}
\begin{tabular}{ m{5cm}  m{3cm} r}
\toprule
\textbf{Research Question} & & \textbf{Total} \\
\midrule
    \multicolumn{2}{l}{RQ1:\RQOne} &  39	\\ 
\hline
    \multirow{3}{*}{\parbox{5cm}{RQ2.1:\RQTwoDotOne}}&{\parbox{3cm}{Software development governance}}&40\\ \cline{2-3}
    &{\parbox{3cm}{Software}} &62\\ \cline{2-3}
    &{\parbox{3cm}{Package and version}}&7\\
    \hline
        \multicolumn{2}{l}{\parbox{8cm}{RQ2.2:\RQTwoDotTwo}} &4	\\ 
\hline
    \multicolumn{2}{l}{\parbox{8cm}{RQ2.3:\RQTwoDotThree}} &16	\\ 
\hline
    \multicolumn{2}{l}{\parbox{8cm}{RQ2.4:\RQTwoDotFour}} &17	\\ 
\bottomrule
\end{tabular}%
 \end{minipage}
\end{table}


\subsection{Concept of Software Trust in Literature}
\label{RQ1}

In this subsection, we focus on answering RQ1: \textit{\RQOne}

As we introduced in the introduction section, the common perception of trust is that it is a belief that the trusted person will actually do what is expected by trustor. In fact, researchers have interpreted trust differently in various domains. In software engineering alone, trust is understood differently. For example, \citet{jadhav2011framework,garcia2015reputation,grodzinsky2011developing} emphasize that trust involves risks, defects, and vulnerabilities; \citet{wang2019updating,nunes2019explaining,garcia2015reputation,ghapanchi2015longitudinal} argue that trust is viewed as a function of quality of service; \citet{jackson2009direct,alarcon2020trust,trvcek2018brief} state that software trust encompasses trust in software producers. Table \ref{Trust def} gives the attribution of software trust we collected from the literature. This table only contains several high frequency attributions, understandings of trust can be found more in the literature.

\begin{table}[!htbp]
\begin{center}
\begin{minipage}{\textwidth}
\caption{Attribution of software trust based on the frequency of metrics in the definitions of software trust from the literature.}
\label{Trust def}
\footnotesize
\begin{tabular}{p{2cm} p{7.5cm} p{1.5cm}}
\toprule
\textbf{Attribution} & \textbf{Definition}   & \textbf{Source} \\ \midrule
Availability& Trust is A capability of delivering requested functionality or services. &\cite{boyes2014trustworthy} \\ \hline
Boundary   & It is important to identify the critical properties and determine what level of confidence is required in addition to software, physical equipment and humans. &\cite{ghebremedhin2012combining,jackson2009direct}  \\ \hline
Calculated, measured, evaluated & Though software cannot be seen or touched, there must be an evaluation and measurement of its trustworthiness in order to ensure the success of the software. &\cite{rosenberg1998software}  \\ \hline
Confidence& Trustworthiness of software is determined by the degree of confidence that exists that it meets certain requirements. &\cite{del2011survey} \\ \hline
Credibility& A comprehensive qualitative description of the software's availability, stability, performance, security, durability, and supportability under specified conditions and specified time. &\cite{hong2011research}   \\ \hline
Dependability& A system is dependable means it can be relied upon or trusted to perform a certain task. In other word, it only be justified if there is evidence that the system has the capability to function without certain failures. &\cite{jackson2009direct}  \\ \hline
Evidence, Observation & Trust refers to an aspect that is based on evidence or observation. Alternatively, it includes subjective aspects of an entity's opinion, such as the presence of human information that indicates that such incidents will not occur. &\cite{jackson2009direct}   \\ \hline
Weakness   & Trust is the ability of a system to perform its required functions despite disruptions brought about by natural or human-made events, hostile attacks by internal or external parties, and/or system errors caused by hardware and software issues. &\cite{jadhav2011framework} \\ \hline
Human, vendor   & One trust is about trust in technological artifacts, while the other is about trusting the vendor’s capability to fulfill the outsourced task and provide quality services.  &\cite{goode2015rethinking} \\ \hline
Reputation   & Reputation complements trust and can be seen as public perception of trust.  &\cite{nunes2019explaining} \\ \hline
Quality   & The notion of trust represents the overall system quality indicator, which should reﬂect the quality of performance and security, reﬂecting their static and dynamic nature.  &\cite{jadhav2011framework} \\ \hline
Structural Assurance   & No matter the specific characteristics of the technology, there are structural conditions such as contracts, guarantees, support, and other safeguards that should guarantee the success of the technology in general. &\cite{guo2014ratings} \\ \hline
Subjectivity  & Software trust depends on the individual's past experiences, their biases, their education, and so forth. It is only through the provision of specific, detailed guidelines that this subjective emphasis can be removed.   &\cite{guo2014ratings} \\ \hline
Temporal, Dynamicity & Software trust is built continuously and gradually, and changed over time as more evidence or experience arrives. More evidence is needed to form a high trust. &\cite{guo2014ratings}  \\ \hline
Transitivity & If A trusts B and B trusts C, then it may be concluded that A trusts C to a certain degree. Transitivity is widely used in trust-based recommender systems and systems authorized by certificate authorities. &\cite{guo2014ratings} \\ \hline
Transparency & A computer is inherently opaque, and one must determine whether the level of transparency provided is sufficient for the system to be trusted. Therefore stakeholders have to communicate the truth, keep their  commitments , and consider the interests of each other. &\cite{mercuri2005trusting} \\ \hline
Uncertainty   & Trust is essential for almost every situation where either uncertainty exists, or undesirable outcomes are possible, meaning that the end-user is exposed to the risk that the software might fail to fulfill its prescribed responsibilities whether intentionally or unintentionally.&\cite{mcknight2011trust}   \\ \hline
Vulnerability   & As a result, end-users accept any vulnerabilities as a part of the software. An alternative understanding is that a system is trustworthy when it functions as intended without unnecessary problems.   &\cite{rosenberg1998software};\cite{jadhav2011framework} \\ \bottomrule
\end{tabular}
\vspace{-184.13041pt}
\end{minipage}
\end{center}
\end{table}

Software trust is inherently predictive based on historical data, regardless of its focus~(\cite{jadhav2011framework}; \cite{kula2015trusting}; \cite{jackson2009direct}; \cite{alarcon2020trust}; \cite{trvcek2018brief}; \cite{mcknight2011trust}). According to this interpretation, we need to pay attention to two aspects: the source of historical data and which aspects will have an impact on trust, which we are basing our judgments and predictions about trust. In Section \ref{Trustsource}, we discuss the source of interaction histories, in Section \ref{trusttypes}, we explain three aspects that require prediction, i.e., software, human and structural assurance (the third column in Figure \ref{fig:trust def}). 
\subsubsection{Source of Software Trust}
\label{Trustsource}

Software trust is established through the software producer's and the software's characteristics, e.g., how the first impressions of the software are, which features of the software are more important, or which are regarded as low priority or irrelevant. Sometimes, end-user first impression may influence the final choice. For instance, while a project with spelling errors in the README.md file does not necessarily indicate poor quality, an observer may conclude that there is insufficient attention given to the project by software producers.

\begin{figure*}[!htbp]
    \centering
    \includegraphics[trim=30 180 30 120,clip,width=\textwidth]{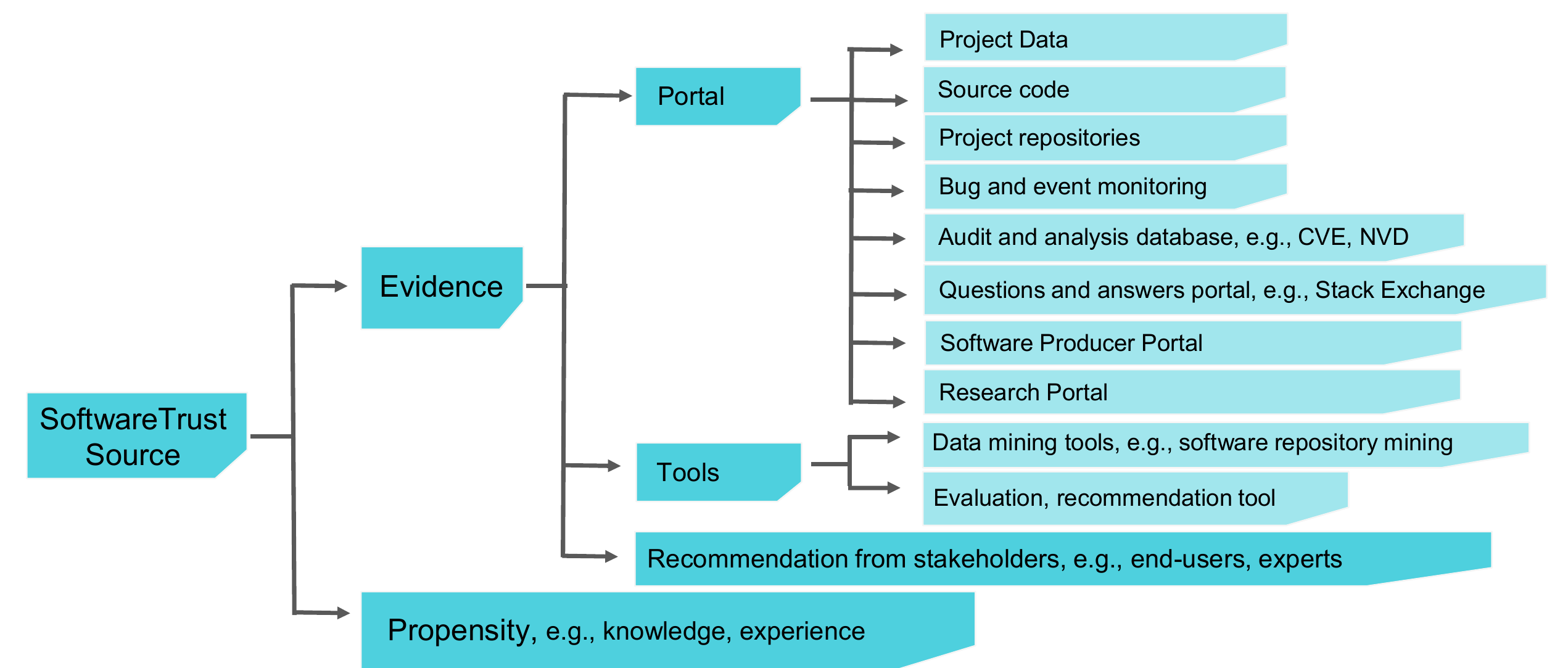}
    \caption{\footnotesize\textit{In this model, we outline the main sources of trust, split into evidence, e.g., portals of software products or software producing organizations, and propensity, e.g., software end-users' knowledge or experience~(\cite{amoroso1991toward}).}}
    \label{fig:trust source}
\end{figure*}

As shown in Figure \ref{fig:trust source}, trust is based on either \textit{evidence} or \textit{propensity}. Evidence includes information from the outside world, such as vulnerability reports from various trusted repositories, for example, Common Vulnerabilities and Exposures database (CVE) and National Vulnerability Database (NVD); portals for software reviews, or communities. However, for proprietary products, since such data is not open to the public, the way for end-users to get relevant information is from the software or organization portals, or from other users' experience with the software. The issue then is that information from the software producers directly could be one-sided and unverified, insufficiently reflecting the reliability of the software product. In contrast, open-source software products provide evidence in the form of project data, source code, and bug reports, that are easily accessible since the repositories, e.g., Github, GitLab, or BitBucket are public and end-users do not require authorization to access them~(\cite{manikas2013software}). End-users can analyze this ``raw data'' to determine the trust of the product easily. Additionally, evidence may come from stakeholders, such as internal experts or external consultants for advice or evaluation, or experienced end-users or experts, such as knowledge sharing or questions and answers from stack exchange or stack overflow. Besides, various analytics tools and recommendation systems are sources of trust. For example, the social network-based recommendation system, which aims to make personalized recommendations by providing users with the most relevant data from friends~(\cite{krishnan2008predicts}).

The other part is the propensity of software end-users, shaped by their experience, education~(\cite{amoroso1991toward}), cognition, and rationality. Rationality stems from the collection of information, nevertheless, there are situations, such as when faced with a high level of uncertainty or a time constraint, in which rationality may not lead to the optimal decisions~(\cite{evans1996rationality}). Similarly, in the presence of factors such as preconceptions, memory, or responsibility, cognition may influence the probability of making an incorrect decision by the end-user~(\cite{chattopadhyay2020tale}). However, cognition is not always wrong~(\cite{decan2018impact}; \cite{jadhav2011framework}). For example, a knowledgeable end-user can determine whether a software package is trustworthy and appropriate for the current project based on his previous experience. Correspondingly, a person without any IT background and experience in using software packages cannot be expected to find a trustworthy one. 
An interesting fact about propensity is that when end-users are gathering information, they tend to be biased and suspicious of positive information, perhaps because producers eliminate negative information on a large scale, causing end-users to distrust them~(\cite{tavakolifard2012taxonomy}).

Preference is another kind of propensity. 
All attributes should be weighed according to stakeholders' value propositions~(\cite{li2012integrated}). A software product may not give the same level of confidence across all functions but still remain dependable~(\cite{jackson2009direct}). For example, a bank system may incorrectly calculate the expiration date of a credit card but must not disclose the customer's information; or a cellphone may be unable to change the ringtone, but it can definitely make a call. Software trust has boundaries, which encompass the specific functionality, timing, user roles, experience, and requirements of the software~(\cite{jackson2009direct}). Defining the boundaries of software trust facilitates the focus of the analysis on specified areas~(\cite{ghebremedhin2012combining}), as well as clarifying the priority requirements and what level of trust. All software trust we discuss should be predicated on the boundaries of software.

\subsubsection{Types of Trust}
\label{trusttypes}

\begin{figure*}[!htbp]
    \centering
    \includegraphics[width=0.9\textwidth]{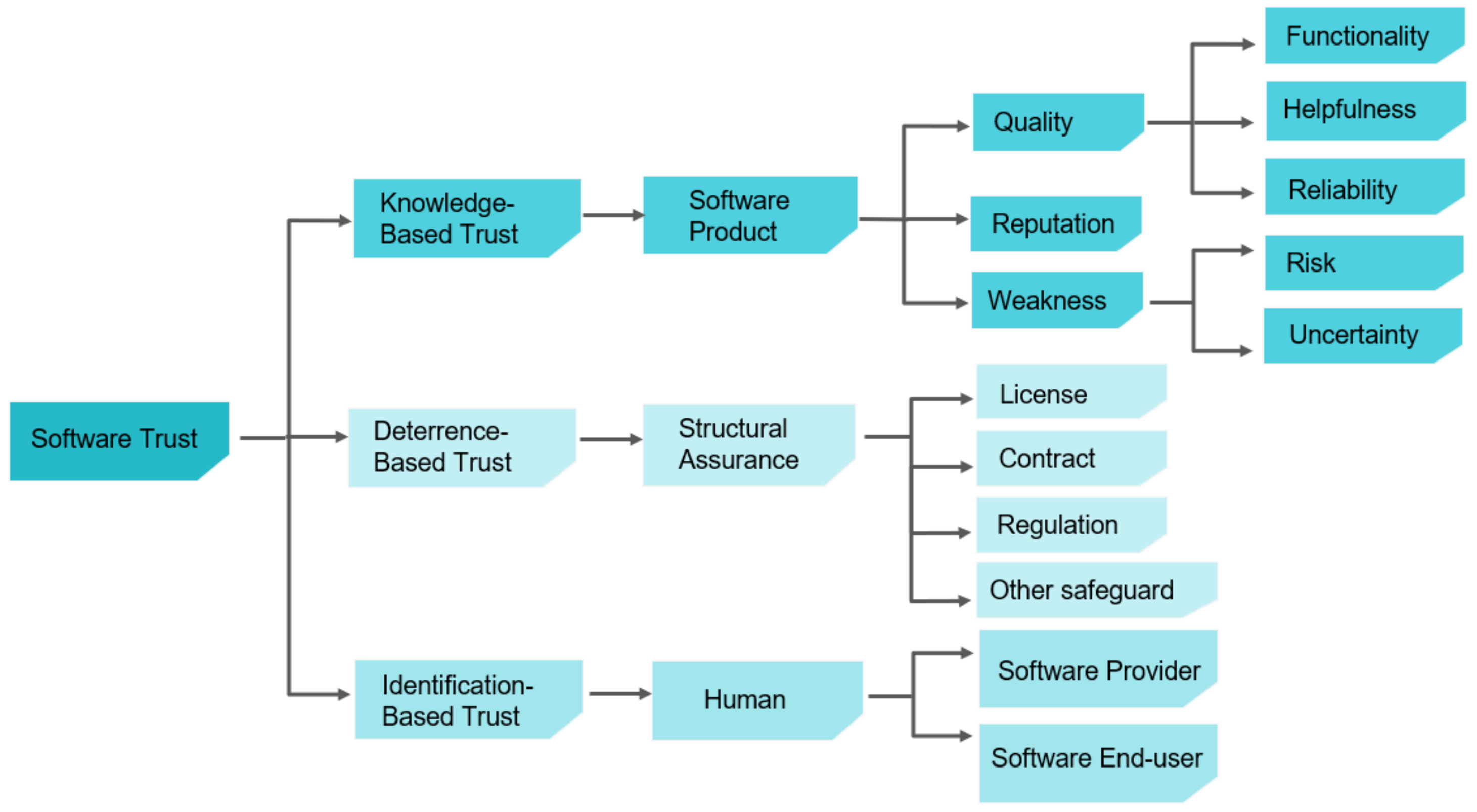}
    \caption{\footnotesize\textit{This model depicts the three types of software trust: knowledge-based trust, deterrence-based trust, and identification-based trust~(\cite{sabherwal1999role}), as well as the corresponding factors that affect software trust, i.e., software products, structural assurance, and human.}}
    \label{fig:trust def}
\end{figure*}

There is a large volume of publications describing trust.~\citet{das2001trust} describe trust as goodwill trust and competence trust;~\citet{sheppard1996network} define trust as deterrence-based trust, knowledge-based trust, and identification-based trust; and ~\citet{sabherwal1999role} specifies trust as calculus-based, knowledge-based, identification-based, and performance-based trust. Till now, there is still no consensus on the definition, as trust is different in different contexts~(\cite{heiskanen2008control}). When extended to software trust, we find that identification-based trust proposed by~\citet{sheppard1996network} coincides with traditional social science trust, which is stakeholder-oriented trust, i.e., trust related to software engineers, organizations, and communities, referred to software producing organizations in this study. Another one, knowledge-based trust, is in line with the trust in software artifacts and products, which is based on knowledge of a technology or product to predict whether it will function as intended~(\cite{jadhav2011framework,kula2015trusting,jackson2009direct,alarcon2020trust,trvcek2018brief,mcknight2011trust}. The subject of this study is software trust in the context of SECO, hence, as shown in the Figure \ref{fig:trust def}, we refer to the classification proposed by~\citet{sheppard1996network}. These three types of trust correspond to trust in software product, structural assurance, and trust in human, respectively. The correspondence is described in the following section. We highlight the major factors as \textbf{boldface} text (the fourth column in Figure \ref{fig:trust def}), sub-factors as \textit{italics} text (the fifth column in Figure \ref{fig:trust def}).

\noindent \textbf{Knowledge-based Trust - Trust in Software}
\label{TrustInSoftware}
\begin{itemize}
\item[]
In Figure \ref{fig:trust def}, trust in software is a kind of knowledge-based trust. Based on the amount of citations, it primarily arises from the quality, reputation, and weakness of software products. 

\textbf{Quality} is crucial to knowledge-based trust. In software evaluation, it refers to the functionality, reliability, and helpfulness~(\cite{mcknight2011trust}). \textit{Functionality} is the most basic condition of a trustworthy software product, \citet{moumane2016usability} defines it as the purpose of the system and what it does. Thus, the software product should provide confidence to the end-users that this product is what the end-users want. According to IEEE, \textit{reliability} refers to the capacity of a system or component to perform specified functions for a specified period of time under specified conditions~(\cite{rosenberg1998software}). Most software development managers equate reliability with correctness, they use frequency of failure, fault tolerance and recoverability to gauge how reliable a software product is~(\cite{berander2005software}). \textit{Helpfulness} refers to the software product providing adequate and responsive guidance to users, for instance, if a help function supports users independently finding their way in the product~(\cite{mcknight2011trust}; \cite{lankton2015technology}). These three are the most mentioned subfactors in the literature, some attributes, such as \textit{security, resilience, and safety}, are also mentioned in the literature~(\cite{jadhav2011framework}; \cite{del2011survey}; \cite{lai2011trust}). Although they are not included in the Figure~\ref{fig:trust def}, we believe that they are critical to ensuring software trust as well. 
 
\textbf{Reputation} is the average belief of the relevant audience that you do or do not possess a particular trait, and it is built gradually over time~(\cite{barclay2015reputation}). Therefore, reputation is believed by the majority of people.~(\citet{nunes2019explaining}) believe ``\textit{reputation complements trust, and can be seen as a public perception of trustworthiness}'', hence, software trust is built on a public foundation, accumulating a reputation for continuous, high-quality performance~(\cite{mailath2001wants}). This performance is not only about the high quality of the software itself, but also encompasses whether the software producer can deliver high-quality services.  
  
\textbf{Weakness} refers to risk or uncertainty, anything predictable or unpredictable that would cause problems or failure of the software or software producer. Some studies have argued that \textit``{trust is the willingness to take risks under uncertainty}''~(\cite{cho2015survey}), it describes ``\textit{the probability of a particular piece of software to elicit a security failure within a certain time frame}''~(\cite{bugiel2011scalable}). Trust constitutes a distributed, flexible concept that improves the decision-making process without the need for heavy management, given the various risks and uncertainties from the software itself or from the software vendor~(\cite{moyano2016model}). These uncertainties can entail requirements, technology, cost, schedule, support resources, project management, and external attacks. We look at separately the relationship between risk and software trust and the relationship between uncertainty and software trust. \textit{Risk} is the situation where the decision-maker knows the possible outcome of the decision and the probability of its occurrence~(\cite{Park2017}). Therefore, risk is predictable. Trust comes with risk, and the greater the risk, the greater the trust required. There is no doubt that there is risk in using software~(\cite{song2010entropy}). Therefore, we should focus on how software producers control risks and how software end-users predict risks and understand how software producers control risks. For example, it is common that expecting zero-defect software is unrealistic, so instead of being concerned about the presence of vulnerabilities in software, end users are more interested in when the vulnerabilities found can be resolved in a timely manner. \textit{Uncertainty} has been mentioned in several studies on software trust. The sources of uncertainty are diverse and can be either the quality of the software or the reputation of the software vendor. Although it is unpredictable~(\cite{Park2017}), a large part of this uncertainty is not caused by the internal problems of the software but by the external environments in which it is deployed. This is because, a part of the uncertainty can be exposed during the software testing phase or in the interaction with the external environment that may produce unexpected disturbances~(\cite{muller2021trust}), timely fixes and remedies can eliminate this part of the uncertainty. In addition, modeling and predicting disruptions or impacts from unknown external environments can be difficult, such as the multi-vector attacks on the software supply chain. Therefore,more uncertainty comes from these external environments. However, the significance of uncertainty for trust is that if there was no uncertainty, all information would be objectively available, and end-users could make judgments independently and with confidence, then trust is not needed.

\end{itemize}

\noindent \textbf{Identification-based Trust - Trust in Human}
\label{TrustInHuman}

\begin{itemize}
\item[]
In the model in Figure \ref{fig:trust def}, trust in human is a type of identification-based trust. The identification-based trust exists because when all stakeholders understand each other's intentions and expectations, a connection is built~(\cite{sheppard1996network}). Considering the connection between software end-users and software producers, \textbf{software producers} need to understand end-users need and their expectations; in turn, \textbf{end-users} need to understand the performance of software producers. That is, knowledge of software producers' intentions and expectations is based on a good understanding of their characteristics, such as prior performance, experience, skills, and team size. Therefore, it can be seen that identification-based trust is built on knowledge-based trust~(\cite{sabherwal1999role}), based on a full understanding between the software producers and end-users. 

Trust in software producers is discussed in the literature by examining whether they are considered professional and reliable and can produce high-quality software products and provide quality services~(\cite{jadhav2011framework}). In addition, how software producers remediate and recover from risks and problems is also considered to be a factor that has an impact on the reputation of software producers as it affects the end-user's willingness to maintain a long-term service relationship with the software producer~(\cite{bennett2000service}). 
\end{itemize}
\noindent \textbf{Deterrence-based Trust - Trust in Structural Assurance}
\label{TrustInssurance}
\begin{itemize}
\item[]
Deterrence-based trust is built by the trustee's fear of the consequences of breaking a rule, law, or contract. That is, there are \textbf{licenses}, \textbf{regulations}, or \textbf{contracts} in place in order to prevent someone from harming another, so the trustor can trust that the trustee will not do anything that could cause her harm~(\cite{sheppard1996network}). Therefore, to ensure that all stakeholders' interests are not compromised, clear premises or assumptions, clear responsibilities to be fulfilled by stakeholders, and contracts or clauses with clear penalties are required. When end-users encounter an unknown software producer or software, the emergence of such assurance protects the expectations and interests of end-users~(\cite{wright2010trust}). Here we call them \textit{structural assurance} (SA), defined as ``\textit{the degree to which consumers believe that institutional structures like guarantees, regulations, promises, legal recourse, or other procedures are in place to promote success}''~(\cite{sha2009types}). 
\end{itemize}
\subsubsection{Definition of Software Trust}
\label{STdef}

Software ecosystems are complex networks of overlapping supply chains of software, data, and services. These supply chains have \textit{upstream flows} from end-users to software engineers, that for instance contain money and data, and \textit{downstream flows}, that for instance contain source code and packages. \textit{\textbf{Software ecosystem trust}} constitutes all upstream and downstream trust that is present between the stakeholders that make up the SECO. This includes, for instance, the trust of end-users in software engineers to build solutions for their (business) problems, and the trust of software engineers in software package management systems to correctly transfer their software to end-users. The concept of software ecosystem trust is relevant, because without software ecosystem trust, it becomes impossible for the SECO to fluently deliver value to its stakeholders. Software trust is different from software ecosystem trust. The definition of software trust is based on the complex relationships between software end-users and producers in the software ecosystem. To define software trust in this context, we refer to the definitions and attributions of trust, and define software trust as follows:

\fbox{%
\parbox{0.95\textwidth}{%
\begin{flushleft}
Software trust refers to the willingness of software ecosystem actors to accept risks based on subjective beliefs. It is essentially an upstream trust in that actors expect assurance that other actors on top of a common technology platform can exhibit reliable behavior and provide them valuable software products. Software products are expected to be able to carry out their intended functions in the presence of uncertainty, and run consistently and reliably without interruption.
\end{flushleft}
}
}\leavevmode\newline


\subsection{Trust Factors of Software Products}
\label{RQ2.1}

This subsection argues \textit{RQ2.1 - \RQTwoDotOne} As discussed in Section \ref{codingscheme}, based on CUT, we classify the trust factors as intrinsic trust factors and extrinsic trust factors. Intrinsic trust factors refer to the features, code, and architecture of the software product, such as software quality, code quality, and known vulnerabilities. And extrinsic trust factors refer to attributes bestowed on the software product by the outside world, such as popularity or reputation, as well as software-based services and auxiliary materials that need to be delivered to the end-users. We highlight the major factors as \textbf{boldface}, as we do in the subsequent sections.

\subsubsection{Intrinsic Trust Factor}

Intrinsic trust factors represent internal physical attributes, rather than being determined by the environment. The following factors are most frequently discussed in the literature:

\noindent\textbf{Security, Vulnerability, and Attack Proneness} - Security is one of the main challenges in software. With the continuous emergence of new types of cyber-attacks, software attacks show a gradual trend of a large-scale, multi-vector, designed to infect multiple components of the SECO. According to Sonatype, in 2021 the world witnessed a 650\% increase in software supply chain attacks, aimed at exploiting weaknesses in upstream open source ecosystems~(\cite{Sonatype2021OSS}). 

Typically, software engineers are not qualified to spot vulnerabilities in code, community knowledge, such as publicly accessible vulnerability databases, can help~(\cite{viertel2019detecting}). Literature shows that there are three sources of vulnerability data; the first is public databases, such as CVE, NVD, or snyk.io; the second is software producers' databases, such as Microsoft security bulletins; the third is specialized databases~(\cite{massacci2010right}). Yet, the quality of the data in the vulnerability database, in particular the timeliness, consistency, and integrity, is an issue that requires immediate attention~(\cite{anwar2021cleaning}).

Vulnerabilities are not addressed in a timely manner. The results of a study of Node Package Manager (npm) showed that most of the reported vulnerabilities were moderately or highly critical, and while the time required to fix them varied by severity, fixing them all took a long time. The probability of a vulnerability being fixed within one month of discovery is 50\%, while the probability of a fix after only 6 months is 74\%. Furthermore, more than half of all dependency packages are affected by vulnerabilities in upstream packages. One interesting aspect is that most of the affected dependency packages are not automatically fixed with the fix of an upstream package. The reasons typically are the lack of proper maintenance of the package or strict dependencies~(\cite{decan2018impact}). However, not all vulnerable packages or libraries will propagate vulnerabilities since not all code in the package or library will actually be used~(\cite{hejderup2018software}). In addition, a number of organizations defense approaches are more oriented towards preventing access to sensitive data or finding bugs in software and software packages than malicious packages~(\cite{Duan2020Measuring}).

Unfortunately, there is still no uniform standard to measure software security~(\cite{wang2015trustie}). This may be because software security is influenced by too many factors, such as environmental factors, complex software supply chains, project management, or the adoption of programming languages. For the software itself, the complexity of the requirements, the complexity of the code, the technology used, and even the skills of the engineers all have an impact on software security, with the problem being that these factors are hard to quantify. Meanwhile, it is unrealistic to request software testers to cover all security requirements perfectly. In addition, different stakeholders have different perceptions of software security. Although organizations and governments tend to place a high priority on software security~(\cite{wang2015trustie}), typically, software engineers do not it put in the first place~(\cite{vargas2020selecting}). In particular, software security requires significant resources to develop, evaluate, and defend, small and medium-sized companies may not be able to afford this investment. 

\noindent \textbf{Quality and Development Process} - Factors regarding quality revolve around the quality characteristics of the software itself and about the activities of the software during its life cycle. Base on ISO/IEC 9126:2001, software quality has been defined as ``the totality of characteristics of a software product that satisfy stated or implied needs''~(\cite{coallier2001software}), which means the software product and process should satisfy (1) specific functional requirements and; (2) end-user needs or expectations~(\cite{miguel2014review}).

There are various well-known software quality models. McCall’s Quality Model proposes 11 factors that describe the external view of the software (user view) and 23 factors that describe the internal view of the software (developer view)~(\cite{al2011software}). Similarly, Boehm’s Quality Model uses a hierarchical quality model structure but focuses on a wider range of characteristics, i.e., as-is utility, maintainability, and portability~(\cite{berander2005software}). FURPS \& FURPS+ are developed by Hewlett-Packard, identifying software quality factors through functional requirements and non-functional requirements~(\cite{khosravi2004quality}). Dromey's Quality Model recognizes that quality is assessed differently for each product and therefore requires a combination of product characteristics and quality attributes to determine how each attribute affects quality attributes and identify software shortcomings~(\cite{berander2005software}). ISO/IEC 9126:2001 incorporates these models, it has four parts (quality model, external and internal metrics, and quality metrics) covering six significant areas of software evaluation: functionality, reliability, efficiency, maintainability, portability, and usability~(\cite{berander2005software}). However, there are no indications on how to quantify these factors and subfactors, and this model may reflect the perception of software engineers instead of end-users~(\cite{challa2011integrated}; \cite{cote2005evolution}). It has been replaced by ISO/IEC 25010, which focuses on the characteristics of software internal quality, software external quality, and software usage quality. The most important point is that ISO/IEC 25010 proposes trust as ``the degree to which a user or other stakeholder has confidence that a product or system will behave as intended'' in quality of use. Several studies have extended based on these well-known models, such as the Fuzzy Software Quality Quantification Tool (FSQQT), it takes as input several real-time values of metrics and uses as output quantified software quality based on end-user, and engineers' perspectives. This study shows that end-users care more about reliability and usability, while engineers focus more on functionality, efficiency, portability, and maintainability~(\cite{challa2011integrated}). This is probably because end-users do not know the software's code, architecture, or efficiency, so they are more interested in the maturity of the software, fault tolerance, and ease of operation and learning.

Table \ref{tab:factor} shows the number of primary factors and subfactors stated in the models of McCall's Model, Boehm’s Quality Model, Dromey's Quality Model, ISO/IEC 9126:2001, ISO/IEC 25010, FURPS \& FURPS+, and FSQQT Model. We found that reliability is the most critical primary factor in quality models, the following are efficiency, portability, and usability, accuracy, testability, and adaptability are the most stated subfactors.

\begin{table}
\centering
\caption{The table shows the primary factors and counts base on McCall's Model, Boehm’s Quality Model, Dromey's Quality Model, ISO/IEC 9126:2001, ISO/IEC 25010, FURPS \& FURPS+, and FSQQT Model.}\label{tab:factor}
\begin{minipage}[t]{0.48\linewidth}
\centering
\vspace{0pt}
\begin{tabular}{lr}
\toprule
\textbf{Primary Factor}                  & \textbf{Total} \\ \midrule
Reliability      & 6              \\ 
Efficiency       & 5              \\ 
Portability      & 5              \\ 
Usability        & 5              \\ 
Functionality                            & 4              \\ 
Maintainability  & 4              \\ 
Correctness      & 2              \\ 
Testability      & 2              \\ 
\bottomrule
\end{tabular}
\end{minipage}
\hspace{\fill}
\begin{minipage}[t]{0.48\textwidth}
\vspace{0pt}
\begin{tabular}{ c c }
\toprule
 \textbf{Subfactor}  & \textbf{Total} \\ \midrule
Accuracy        & 5              \\ 
Testability     & 4              \\ 
Adaptability    & 4              \\ 
Interoperability  & 3 \\ 
Security                             & 3              \\ 
Maturity        & 3              \\ 
Fault tolerance & 3              \\ 
Operability     & 3              \\ 
Installability                          & 3              \\ 
Replaceability                          & 3              \\ 
Reusability                             & 3              \\ 
Consistency                             & 3              \\ 
Completeness    & 2              \\ 
 \bottomrule
\end{tabular}
\end{minipage}
\end{table}

Software quality is not accidental; it is achieved through management and must be assured throughout the entire product life cycle. Reliability is an attribute of quality, the quality attributes applied at each stage of the SDLC determine the software reliability, the focus is on error prevention, especially the early stages of the lifecycle~(\cite{rosenberg1998software}). The advantage of the OSS project is that the project data is public, so there are ``many eyeballs'' to help find and fix defects~(\cite{mohagheghi2007quality}), thus ensuring more credible software quality.

It is suggested by models or standards that evidence of software quality or software trust should cover the information obtained at each stage of the development process. For instance, IEEE Std 982.2-1988 indicates the relationship between software reliability and each life cycle phase, including concept, requirement, design, implementation, test, installation and checkout, operation and maintenance, and retirement. In spite of the fact that this standard was published in 1988, it does, to some extent, reflect the information at each stage of the software development life cycle that influences the measurement of quality. Additionally, a number of Capability Maturity Models (CMM) address quality issues from five maturity levels of software development processes, i.e., initial level, repeatable level, defined level, managed level, and optimizing level~(\cite{ellison2016measuring}; \cite{berander2005software}). Repeatable level focuses on project management, the defined level focuses on engineering process, managed level focuses on product and process quality, and the optimizing level focuses on continuous improvement. \citet{he2009reference} argue that trusted components is affected by the development process, therefore, the trustworthy proof should be collected through three phases, i.e., development phase, submission phase, application phase. Submissions and applications are primarily concerned with the quality factors in the quality models, for instance, submission phase includes the proof of functionality, reliability, and usability, application phase emphasizes the proof of effectivity, security, and efficiency. Similiarly, \citet{wang2011trustie} states that trustworthiness evidence includes three aspects, development-stage evidences, delivery-stage evidences, and application-stage evidences.

Additionally, with the introduction of third-party components, their features cannot be assessed using traditional quality attributes. Additional concepts of software quality arise, such as configurability, customizability, reusability, scalability, availability, trackability, and compatibility to the quality requirements for software products produced by traditional software development processes~(\cite{miguel2014review}; \cite{chau1994selection}; \cite{challa2011integrated}). These concepts emphasize the need for third-party components to be flexible in design and application, easy to reuse, change, and extend. The object is not only making the third-party components reusable, but also making them adapt components to end-users' systems and requirements~(\cite{wang2019updating}). 

\noindent \textbf{Source Code Quality and Architecture} - Several researchers calculate software quality as the ratio of the total number of defects identified in a certain period of time to the total number of code lines~(\cite{liu2007design}). Thus, it is not hard to find that software quality, especially for open-source software, depends on the quality source code~(\cite{ghebremedhin2012combining}) and its attributes, for instance, availability, understandability, completeness, conciseness, portability, consistency, maintainability, testability, usability, reliability, structuredness, and efficiency~(\cite{crowston2003defining}; \cite{jadhav2009evaluating}; \cite{wheeler2011free}). The most frequently stated impact aspect concerning source code in the literature is availability. Reasons abound for this, one of which may be that end-users or downstream engineers need to inspect or analyze source code in order to determine code quality as well as software quality.

Programming languages may not directly affect software trust. Still, the experts believed that 68\% percent of the programming language features, for instance, application domain, platforms, maturity level, and comprehensiveness, have impacts on the aspects of usability, cost, stability, ownership, guarantee, reputation, support, and maintainability~(\cite{farshidi2021decision}). Projects written in more popular programming languages, are more successful in market penetration and attracting human resources~(\cite{ghapanchi2015longitudinal}), which indirectly gives end-users greater confidence to choose their software. In addition, several aspects need to be considered: (1) the project itself, for instance, as a back-end server, it is not possible to consider a low performance interpreted language like Python; (2) support of programming languages; whether in-house developers are proficient in the language; (3) whether subsequent software packages need to be purchased or used, and the interfaces. 

We try to collect a set of indicators for source code analysis, result can be found in \href{https://data.mendeley.com/api/datasets/xn4jk93g4t/draft/files/b36ddc93-0f42-4227-b966-4c07d2a7e429?a=9c51b45f-c943-4a22-814f-5baf59fb2896}{this link}, which are also the most commonly used fields in various source code analysis tools. Examples include STRAM (Security, Trust, Resilience, and Agility Metrics), a system-level trustworthiness metric framework~(\cite{cho2019stram}), and a framework proposed by Jadhav and Sonar for the evaluation and selection of software packages ~(\cite{jadhav2011framework}), and an updating model based on Meyer's ABCDE(Acceptance, Behavior, Constraints, Design, Extension) model of  software component trustworthiness per users feedback~(\cite{wang2019updating}). In forming this set of indicators, we considered, as much as possible, both the goal of the analysis (i.e., to be able to describe the quality aspects of the software process and the resulting product) and the availability of metrics in the data~(\cite{koch2008exploring}), assessing process and product evidence ~(\cite{donohue2005my}), and product metrics~(\cite{gopalakrishna2005vulnerability}).

The architecture comprises \textit{``a collection of software and system components, connections, and constraints; a collection of system stakeholders’ need statements; and a rationale that demonstrates that the components, connections, and constraints define a system that, if implemented, would satisfy the collection of system stakeholders’ need statements''}~(\cite{gacek1995definition}). It holds a great deal of importance to ensure software quality. It is reported that\textit{``engineers will dismiss an unfamiliar framework and library if they cannot make it work within an hour''}~(\cite{haenni2014quantitative}; \cite{haenni2013categorizing}). Sometimes, end-users just select a package or a library to implement a single function instead of a large system or application, for instance, for brownfield development, the existing software architecture of the project imposes constraints on the selection of the subsequent software package or library, the selected package or library must fit into the existing technology stack of the project~(\cite{vargas2020selecting}). In addition, as software packages are hosted off-premises, they are difficult to integrate with other systems or software packages. End-users and software producers should pay additional attention to the architecture's integration, scalability, and reliability. Software should have the ability to remain available to users within a given time window and maintain a reasonable response time for users even during peak periods~(\cite{godse2009approach}). 

\noindent \textbf{Versions and Dependencies} - The distribution, installation, and updating of software packages or components also have an impact on the integrity and reliability of the software. Distribution requires that the package or component is in the format provided by the software producer, does not contain any malformations or corruptions, and that the installation and updates must also be in compliance with the requirements set forth by the software producer~(\cite{catuogno2017secure}). However, dependencies complicate the software components distribution. Dependencies connect various system components, thus forming a highly interconnected ecosystem~(\cite{hejderup2018software}). However, the version of the package it depends on is specific, hence, a critical activity that must be carried out at each stage of the SDLC is ensuring the integrity and coherence of dependencies with deployment tasks, as a failure in this process can result in disastrous consequences~(\cite{catuogno2017secure}). 

Considerable security vulnerabilities are caused by aging libraries~(\cite{kula2017modeling}), however, it does not mean a new version does not guarantee bug-free software. In addition, the most recent release may break dependencies and result in incompatibility~(\cite{decan2016github}). Thus, considerable organizations choose the stable version over the latest version, they prohibit updates in case of dependency updates~(\cite{pashchenko2020qualitative}). The dependency hell occurs when several packages have dependencies on the same package or library, relying on different incompatible versions. Dependency hell can cause a lot of trouble for installation and subsequent maintenance, and can create security risks~(\cite{abate2011mpm}). Still, numerous software engineers are unwilling to update to newer versions, they downgrade package versions to avoid dependencies and conflicts introduced by the latest versions~(\cite{chinthanet2021makes}), even at the expense of some software security, because they are overwhelmed by the burden of maintaining dependencies~(\cite{GKORTZIS2021110653}). Therefore, the requirements for a package or component are: (1) it has to be authentic; (2) unmodified; (3) fresh; (4) all required dependencies are properly satisfied and ensure dependency, integrity, and consistency; (5) unauthorized installation is prohibited~(\cite{catuogno2017secure}).

\subsubsection{Extrinsic Trust Factors}
\label{exsoftware}

Extrinsic trust factors refer to the external environment characteristics that influence software selection. Although the majority of articles discuss intrinsic trust factors, numerous studies have concluded that end-users prioritize extrinsic trust factors during the software selection process. In comparison to intrinsic trust factors, which lack quantitative indicators or detailed data, extrinsic trust factors have a plethora of data sources on openly accessible portals. 

\noindent \textbf{Structural Assurance} - In Section \ref{RQ1} we underlined that structural assurance is believed to increase software trust. It includes licenses (e.g., GNU General Public License 2.0, and Apache License 2.0), contracts, regulations, or other safeguards (e.g., package promises and legal considerations)~(\cite{guo2014ratings}). End-users may perceive that they are using software in a secure, protected environment as a result of compliance, policies or laws. Contracts define the rights and obligations of the software product, including what each stakeholder expects to accomplish, the feasibility of the goals and respective roles, the actions that must be taken, as well as the responsibilities of the software producer and the solutions to anticipated risks or problems. License is the most important type of structural assurance, it is based on copyright law and confers certain exclusive rights on the software producer of a work, including the right to use, reproduce, modify, merge, publish, distribute, sub-licenses, and sell copies~(\cite{alspaugh2009role}). This means that the license restricts the end-user's ability to use the software. For instance, the General Public License (GPL) specifies what end users may do with the software, such as modify the program and distribute the modified version, which protects not only the copyright of software producers but also the interests of both software producers and end-users, thereby reducing the perceived risk between end-users. Hence, it is reported that end-users are more likely to trust a licensed product~(\cite{osterloh2016trust}; \cite{chinthanet2021makes}). Besides, different versions of the same software may have different license, which end-users should be aware of during the software selection.

\noindent \textbf{Documentation} - Documentation provides the first impression to end-users of a product's quality~(\cite{vargas2020selecting}). It includes the official project documentation, e.g., requirement descriptions, code comments, or testing results~(\cite{national2007software}), and broader forms of documentation, e.g., documentation translation, user comments, or discussion in the forums~(\cite{ayala2011selection}; \cite{chau1994selection}). A survey of npm developers initiated by \citet{chinthanet2021makes} revealed that 73\% of developers believe that the documentation is highly relevant to package quality assessment, with the existence of a run example, and the existence of the README.md file in the root directory of the package considering the top two documentation features. Similarly, as part of our previous study, we interviewed 12 participants, including software engineers, DevOps, architects, and other experts, and ten of them considered documentation crucial to package selection. Typically, they examine data structures, the functionality of features, and the package's compatibility with the project via documentation, one of them confirmed that packages requiring 2 days of reverse engineering due to unclear documentation would be eliminated~(\cite{jansen2021trustseco}).

However, there is a problem with both official documentation and broader forms of documentation. In terms of official documentation, not all types of documentation are available to the public due to software producers do not share project data and source code for free commercial software. In terms of broader forms of documentation, although the documentation for open-source software or packages is free and open, much of the technical information and discussion occurs on mailing lists or forums, making it difficult to collect and maintain valid information, let alone version control of the documentation.

\noindent \textbf{Popularity} - Popularity is defined as the perceived frequency of use of a tool among members of the community ~(\cite{xie2021popularity}). It represents a software product's popularity and is the most straightforward way to determine how widely used and popular the product is in the public. Popularity can be measured as the number of likes or downloads of the software product, and the number of subscribers or views of the information page. Additionally, the number of dependencies reflects the popularity of the software, as a component or library, with more dependencies means more people are using it and it is more stable~(\cite{vargas2020selecting}). 

\noindent \textbf{Reputation} - Reputation is an estimation of the nature and values of a person, a group, an organization, an object, an event, or an activity that is widely shared in a group~(\cite{watson2005reputation}). It is a judgment about the quality of a product and future behavior~(\cite{farooq2016multi}). Reputation can be at the software level or human level. Human-level reputation is discussed in Section \ref{RQ2.3}, The focus here is on the reputation of software. The most direct reputation indicator is the average software product ratings~(\cite{alarcon2020trust}). The reputation of a software product may include the number of releases or versions, product history, and open/closed issues(\cite{alarcon2020trust}), programming language used, clarity of comments, number of bugs reported, tests performed, number of versions, and efficacy~(\cite{seigneur2011trust}) from users who used the software or stakeholders who participated in the development or testing. Reputation has a positive impact on software trust. A good reputation represents a good rating by a certain size of users over a given period of time, and it represents a lower risk of choosing the software. Therefore, in similar software quality, the end-user will choose the software with a higher reputation. However, almost all reputation information is based on a web source, and one of the significant issues with single-source reputation is the vulnerability to falsified information~(\cite{farooq2016multi}), which means that software producers are likely to falsify information or provide only one-sided information to achieve a higher reputation. Besides, numerous proprietary product licenses contain clauses prohibiting public criticism of the product without the software producers' prior permission~(\cite{wheeler2011free}). As a result, it becomes more difficult for end-users to obtain real feedback from other users, making it difficult to be able to assess the trustworthiness of proprietary products difficult.

\noindent \textbf{Cost Factors} - Although it is widely accepted that cost is associated with software selection because solution selection frequently involves a trade-off between quality and cost~(\cite{limam2010assessing}), they do not believe that cost is associated with software trust. This is most evident in the growing awareness of software, particularly open-source software, across a range of industries. However, for several specialized industries, such as banking, they resisted adopting free/open source software despite the high cost of customizing and maintaining the software. This is because they are concerned about the lack of support in the event of a failure or additional damage to their core systems due to a security breach. Even if they adopt free/open source software, it is more likely that they will use it for visualization than for core banking modules. 

Many IT companies and open-source platforms currently use Return on investment (ROI) to evaluate related cost factors, divided into cost of investment, cost of reuse assets, and rework savings. The cost of the investment includes defense cost, which refers to the cost of deploying various alternative solutions; decision cost, which refers to the administrative delays, complexity, and effort incurred in the decision process; recovery cost, which refers to the costs associated with the recovery process to back to a normal operating system state; and defense costs which refers to the cost of defending the system against attackers~(\cite{cho2019stram}). Defense cost is the critical one, it includes, for instance, license cost, training cost, installation and implementation cost, maintenance cost, mitigation cost, and upgrading cost~(\cite{jadhav2009evaluating}; \cite{jadhav2011framework}). Cost of reuse assets and rework savings easily reflect whether the software is trustworthy or not, they measure the cost savings when adopting a trusted library or package, as well as the rework cost savings due to reduced system errors~(\cite{mohagheghi2007quality}). Usually, the two factors are as difficult to gather before the end-users select a software or component. The possibility is that they are predicted or calculated after the software has been in use for a time.

In addition to the financial cost, the time cost and potential cost risks that can impact the cost also need to be considered~(\cite{vargas2020selecting}). For instance, the software will decommission in one year or if the supporting team will be available in a long time.

\begin{figure*}[!htbp]
    \centering
    \includegraphics[trim=110 0 110 0,clip,width=0.9\textwidth]{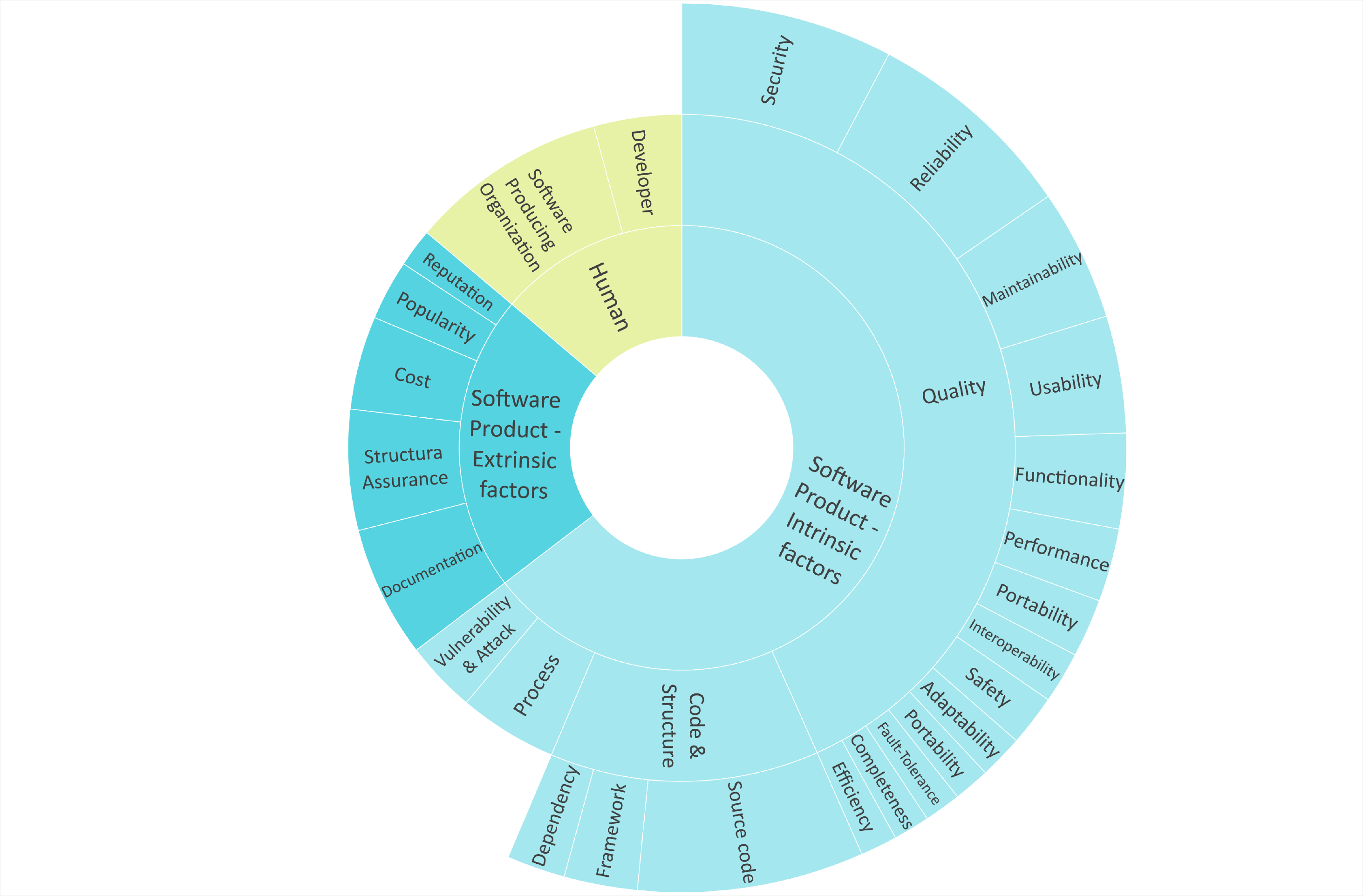}
    \caption{\footnotesize\textit{This sunburst chart shows the major impact factors for RQ2 in hierarchies. Product segments are intrinsic trust factors, Software Product, Human, Structural Assurance, and Cost are extrinsic trust factors. The size of each segment in this sunburst chart indicates the frequency with which this factor appears in the 112 articles selected.}}
    \label{fig:factors}
\end{figure*}

\subsection{Trust Factors of Software Package Managers}
\label{RQ2.2}

This subsection focuses on addressing RQ2.2: \textit{\RQTwoDotTwo}

A package manager is a software tool used to search, install, track, upgrade, and delete packages on the operating system. It automatically connects to a repository and downloads the requested and additional packages needed to build the client application~(\cite{Cappos2008alook}). During the study, we found the current research on software packages is inadequate. Of the 112 articles, only four are about package managers, explicitly refer to Table \ref{papers by rq} for details. 
These articles on the impact of software trust focus on three aspects: (1) dependency hell; (2) Security vulnerability; (3) Vulnerability Detection in the package manager. All three aspects are related to the design of the package manager and the vulnerabilities of the package manager, they are all intrinsic trust factors.

\noindent \textbf{Dependency Hell} - As previously stated, installations are currently experiencing dependency hell, end-users become trapped in a dead-end loop that state-of-the-art package managers are unable to resolve~(\cite{abate2011mpm}). Several researchers argue that a trusted package manager should avoid the dependency hell by selecting the right package sources and versions, especially guaranteeing confidentiality, integrity, freshness, and authenticity of the package and dependency~(\cite{catuogno2017secure}). Currently, attempts have been made, for instance, Nix, a Linux distribution, installs software packages in a purely functional way to avoid the dependency hell associated with the traditional upgrading of /usr/bin/perl(\cite{dolstra2010nixos}). 

\noindent \textbf{Security Vulnerability} - Security vulnerability includes two aspects, bugs and attacks on the metadata and on the repository. In terms of metadata, as explained Section \ref{sec:background}, metadata includes the version of a package and a set of dependencies. An attack on metadata may result in an attacker changing the package's metadata to indicate that it depends on a package containing malicious code formulated by the attacker, or the attacker, introducing unsolvable dependencies or a large number of virtual dependencies that make the installation impossible~(\cite{ayala2011selection}). In terms of the repository, if an online repository contains malicious code when downloading and upgrading packages, it will be imported into the end-user's computer through transitive dependencies~(\cite{hejderup2018software}). For example, GitHub, an online service based on the code version control system, experienced a malware attack in 2020. The malware identifies the NetBeans project file, copies the malicious payload to the project file, executes the code once the NetBeans project is built~(\cite{gonzalez2021anomalicious}). Once the GitHub repository is successfully compromised, the cybercriminals behind the malware have full control over the target open-source project, ensuring that any new build project cannot replace the infected project, and malicious build artifacts remain the same. 

\noindent \textbf{Vulnerability Detection} - The major problem of the package manager design is the lack of vulnerability detection and protection mechanisms. Vulnerability detection tools are generally based on known vulnerability disclosure scanning software components to find vulnerabilities~(\cite{kaminsky2002hack}). However, those tools do not work efficiently, detecting a limited number of possible errors based on predefined rules~(\cite{russell2018automated}). In addition, they also lack oversight of the packages, as well as a mechanism to undo malicious packages after they were installed. For example, when an attacker wants to prevent the package manager, APT or YUM, updating a package from another repository, he refuses to send data or sends data at a very slow rate to the client's open connection, but the package manager lacks monitoring and screening of such situations, does not have any output to indicate problems or forcibly closes the connection ~(\cite{ayala2011selection}).

\subsection{Trust Factors of Software Producing Organizations}
\label{RQ2.3}

This subsection focuses on addressing \textit{RQ2.3. \RQTwoDotThree}

If a product is managed by an organization, it will be more popular~(\cite{jarczyk2014github}, as it gives the end-users the confidence that this product will be produced and maintained by professional staff. However, it is often difficult to determine if an organization is a good self-promoter or a truly reputable organization. Next, we discuss the role that software producing organizations (SPOs) play in software trust assessment.


\noindent \textbf{Reputation} - Reputation means a publicly recognized judgment about the character or status of a person or thing~(\cite{immonen2007trustworthiness}). It is an asset for software producing organizations and gives the organization a competitive edge over others~(\cite{pollock2007technology}; \cite{hoxmeier2000software}). 
Studies have shown that software quality has a positive relationship with an SPO's reputation~(\cite{cai2016reputation}), while an organization's reputation depends not only on the quality of the products it produces, but also on other aspects~(\cite{vargas2020selecting}), for instance, the services they provide~(\cite{jadhav2011framework}), end-user's feedback, and their rating values~(\cite{seigneur2011trust}), positive and timely response, compliance with contracts, and length of experience~(\cite{chau1994selection}; \cite{vargas2020selecting}; \cite{jadhav2009evaluating}; \cite{challa2011integrated}; \cite{capra2011firms}). Length of experience, from the point of view of the SPO, means that the company or organization has experience in product manufacturing, project management, business operations, industry, and experienced in-house experts and engineers~(\cite{chau1994selection}). This extensive experience results in the company or organization developing higher quality and trustworthy software, especially when risks and uncertainties come. They can deal with problems or avoid them more effectively, increasing end-user confidence in their products and services.

\noindent \textbf{SPO's Capability} - SPO's capability is the degree of proficiency in adapting skills to a wide range of activities ~(\cite{goles2008information}). It arises from different factors, including technology skills, firm’s strategy, maturity, and scale. Technology is undoubtedly the capital of enterprises to achieve success and is the core competence. 

Technology skills emphasize technical knowledge in software project management, including programming and system development knowledge, R\&D capability, training, and support. Trustworthy software cannot be developed without programming and system development knowledge, which is the physical guarantee of software production~(\cite{lai2011trust}; \cite{cho2019stram}; \cite{sarrab2014empirical}). This means that the organization must have qualified employees who are proficient in programming, system analysis and design, project management, and other soft skills in order to deliver software on time, on quality, and on schedule, and to effectively control and resolve all risks and issues in the SDLC~(\cite{amoroso1991toward}). It is however not directly measurable, we need to examine the software products they produce and the services they provide instead. 
R\&D capability shows whether the company can develop and maintain current projects and has the potential and innovation to develop new products. It keeps the organization ahead of the competition by continuously innovating and introducing new products and services or improving its existing offerings.  End-users prioritize getting training and guidance in the assistance process~(\cite{vargas2020selecting}; \cite{jadhav2011framework}; \cite{del2011survey}; \cite{sarrab2014empirical}), which include: (1) training and documentation, for example, user manual, tutorials, troubleshooting guide, and training; (2) maintenance and up-gradation, such as technical support and consultancy, communication with end-users, on-site demo and free-trial version, rapid responsiveness, business skills~(\cite{jadhav2009evaluating}; \cite{jadhav2011framework}), management, and operation~(\cite{hunter2013rise}); (3) social network-continued activity; (4) knowledge-sharing user-base~(\cite{wang2015comparative}), e.g., Discussion forums, signs of a functional community~(\cite{10.5555/1593439}). Especially for OSS, a guaranteed, long-term technical support is an important indicator for selecting an OSS software~(\cite{alarcon2020trust}). 

Current software products rely to some extent on third-party components, or many large IT companies outsource all or part of their projects to third-party companies, hence, SPO's technology skills should consider the skills to manage suppliers. For instance, how to select third-part components, how to manage third-part producers, or how to maintain the sourcing strategy. A firm’s strategy includes the management and strategy of the company, the organizations’ culture, and the type of industry. Most organizations and end-users consider influencing the selection process, but they are not the major factors~(\cite{vargas2020selecting}). 

Organizational maturity shows that SPOs have the capability to manage the product well, which is essential to ensure product quality. It is abstract, including, for instance, content management systems, partnership models, rules, and regulations~(\cite{jansen2014measuring}). Although there is no consensus on how to measure it, from the perspective of software end-users, formal procedures and standard security policies will help them evaluate and select software products that are suited for them~(\cite{vargas2020selecting}). 

The scale of the organization is presented by, such as how many people write the code, report problems, and repair defects~(\cite{mockus2002two}), number of members contributing~(\cite{bogart2016break}), and average quality of members~(\cite{challa2011integrated}). The number of engineers working on a project is the most frequently cited factor~(\cite{crowston2003defining}; \cite{jansen2014measuring}; \cite{sen2012open}; \cite{mcclean2020social}; \cite{midha2012factors}; \cite{koch2008exploring}; \cite{scacchi2007free}). These studies show that engineers positively impact the success of a project, especially for open source projects. In particular, by modeling both the number of engineers and subscribers, the results suggest that more engineers in a project may increase the project activity and thus increase the number of subscribers. Similarly, an increase in subscribers may affect the number of engineers by attracting new engineers and contributing new engineers from the subscriber base~(\cite{sen2012open}). Hence, along with the program activity, the number of subscribers, the engineer base is considered by sources of trust information as an indicator of the success of the OSS project~(\cite{subramaniam2009determinants}), as it is obvious that they will increase the popularity of the project. Most studies do not explicitly categorize the engineering base as an attribute of software trust. However, it is often mentioned in studies measuring project success and studies of software trust regarding organizations and communities.

\subsection{Trust Factors of Engineers}
\label{RQ2.4}

This subsection answers RQ2.4. \RQTwoDotFour.

17 selected studies discuss how engineers influence software trust. The identified impact factors are extrinsic trust factors, which are focusing on the following points.

\noindent \textbf{Engineer Abilities and Skills} - Engineer abilities and skills are also related to software trust. Engineering abilities and skills refer to the knowledge and abilities that an individual engineer or a group of engineers should possess throughout the product's lifecycle, they stem from a good working knowledge and experience of different types of operating systems and the ability to learn from others~(\cite{sarrab2014empirical}) These skills and abilities are used to develop and integrate new and stable modules that meet end-user requirements, which includes, for instance, deploying and maintaining software, monitoring news posts and messages in user forums and mailing lists on a regular basis and promptly responding to bug reports~(\cite{norris2004mission}). Several researchers believe that they have a significant impact on the quality of the product they create~(\cite{mcknight2011trust,norris2004mission}; \cite{sarrab2014empirical}; \cite{donohue2003modeling}). Engineer responsiveness, in particular, is viewed as critical regardless of whether end-users trust the software or its output~(\cite{gefen1996developer}). If software engineers are unable to respond promptly to end-users, they will be viewed negatively by end-users and their products, resulting in distrust. Despite the fact that the response time is contingent on the availability of engineers in the project and the priority of end-user problems~(\cite{mockus2002two}). In most cases, end-users are unconcerned with the severity of defects, what matters is that they can be resolved in a short time. In this context, their soft skills are particularly important.

Numerous researchers disagree on the importance of end-users knowing who developed the software. According to a set of interviews with 75 engineers that had been conducted in~(\cite{haenni2013categorizing}; \cite{haenni2014quantitative}), when the end-users select a component or library, they are not concerned with the engineer's identity but with the code's quality and level of maintenance, e.g., the engineer's responsibilities and their response time. The opposite opinion is that the names and numbers of people working on code development and testing must be public, as reputable engineers contribute to the code being more reliable and less defective~(\cite{madanmohan2004notice}). Both views imply that how engineers maintain the project is crucial for software selection.

\noindent \textbf{Engineers' Satisfaction and Happiness} - Besides that, several researchers believe that engineers' satisfaction and happiness with working significantly affect the product quality of their products and the trust that end-users place in them~(\cite{crowston2003defining}; \cite{scacchi2007free}). Engineer satisfaction and happiness have a positive effect on software quality~(\cite{crowston2003defining}). The most obvious benefit is increased productivity, they will stay on tasks longer and concentrate more intently, resulting in significantly more efficient coding and problem-solving~(\cite{graziotin2018happens}). As a result, both the software's quality and the project's success will be significantly enhanced. Engineer satisfaction and happiness are highly subjective factors because each individual's satisfaction and happiness are unique, they cannot be collected and analyzed via specific documents or fields, as test results are. The feasible method is to conduct sentiment analysis on engineers' feelings, for example, through their comments on project portals or websites, to ascertain their level of satisfaction or happiness.

\section{Discussion}
\label{sec:dission}

\subsection{Impact on the Software Selection}

The factors discussed in this study may have a positive or negative effect on the software selection outcome. No factor, on the other hand, can have an absolute positive or negative effect. For instance, vulnerabilities and attacks appear to be detrimental to software trust, as their presence has a significant impact on the quality and credibility of software~(\cite{cho2019stram}) and the reputations of software producers~(\cite{vargas2020selecting}). With the advancement of attack techniques and the expansion of scale and scope, no software can now claim to be secure enough on its own. However, if the software can employ an effective defense to ensure a lower number of attacks over time, and if it is capable of employing more rapid means of resolving the situation following an attack, we can assert that although vulnerabilities and attacks undermine the software, their effect on selection and trust is actually positive~(\cite{gefen1996developer}). Similarly, factors, such as the software's quality, reputation, or the skills of the software producer can have a positive or negative effect, i.e., if the software is of high quality, has a positive reputation, and is provided by a competent provider, they will have a beneficial effect on the software, and vice versa.

The negative or positive impact is determined by how the end-user perceives those factors for a variety of reasons. For instance, several end-users choose low-cost software, whereas several industries, such as banking, in order to get enough resources to develop and maintain the software, choose those with high costs.

There is one situation in which the impact is neither positive nor negative: end-user propensity, as it affects the end-user to conduct the trust process professionally and rationally. The resulting effect on decision-making should be beneficial. However, if the technology of the software evaluation exceeds, such as end-user experience, it may have no effect on decision-making~(\cite{moyano2016model}).

\subsection{Validity Consideration}
We considered validity from the following points. First, our selection of articles was inevitably subjective due to bias and understanding of the articles and several concepts, especially the large number of different similar names for the same concept in the literature, so we could not include all relevant studies included in this SLR. Second, most of the studies and models that we know focus on software quality. There is relatively little research on other hubs in the software ecosystem, stakeholders, and their relationships. The search results are more relevant to FOSS-based discussions because FOSS provides public access to technical data, whereas proprietaries have restricted access. Hence, the analysis of the results may not cover proprietary SECO. Additionally, a considerable number of studies are based on interviews or questionnaires, which may cause the results they present to be not generalized. Moreover, most models are not yet widely adopted in the the open source ecosystem, which means that although the results of several current studies and models come from questionnaires or interviews or are further validated by them, they still lack validation and application in real-world projects, which also further suggests that the results based on questionnaires or interviews may be problematic in terms of generalizability. Finally, we categorized software trust as intrinsic trust factor and extrinsic trust factor, and several subfactors, while this allows for a clarified and focused presentation of the results, we may have inevitably removed several factors that could not be categorized and may have overlooked several specific scenarios that end-users trust.

Next, we will further discuss the role and severity of these factors in the SECO with different stakeholders and experts to ensure that our understanding of these factors is correct and will continue to focus on and supplement with additional impact factors.

\subsection{Lessons and Research Challenges}

Software trust is comprehensive and its factors cover all actors in the software ecosystem. In this subsection, we identify several challenges that make the research more complicated. The first complicating factor is that end-user's propensity to trust is subjective, and even when different end-users are given the same information, the trust they place will be different, and this trust can change over time, the judgment depends on the information they access. Therefore, it is impossible to assign a fixed priority or weight to the factors. 

Second, it is difficult, if not impossible, to find a comprehensive set of models to measure software trust objectively and fairly. The measurement and interpretation of software trust must be dynamic. Third, the interpretation of trust factors should be multi-faceted. It may be difficult to obtain or apply project data as the factor determining trust in closed software, but these data can be used to calculate trust in open source software.

The digitization of society has provided us many affordances and the penetration of IT in it should be considered an unprecedented development in history. However, we are standing at the advent of a new period in history, where societal trust in IT is dropping at an alarming rate. We can say the hay-day of society's willingness to adopt new IT solutions and trust the older ones are over~(\cite{hayes2020algorithms}). With this problem statement in mind, we have set ourselves a research challenge.

The ambitious goal that we have set is to raise the accessibility, reliability, and use of trust data in the worldwide software ecosystem~(\cite{hou2021trustseco}). We envision a future where software can be rapidly and even automatically assessed in a fair and equal manner, using a shared understanding of trust that is collaboratively created and maintained. However, before creating such a free and common infrastructure, several challenges need to be tackled. We define a set of research challenges for the future in \textbf{bold} typeface.

The first challenge is that we are collecting and sharing software engineer data, i.e., data that could reveal their identities and break their privacy~(\cite{van2020core}). This would directly contradict the goals of this project, so we need to find mechanisms for \textbf{protecting personal data long ethical and legal boundaries}, such as the General Data Protection Regulation (GDRP). 

There is an inherent problem with sharing information with a community, as actors could misjudge or misuse the information. It must be noted that \textbf{observed vulnerabilities are not a sign of untrustworthy software} and that if these vulnerabilities are rapidly identified and eliminated, they actually might be a sign of trustworthy software. Furthermore, we need to ensure that \textbf{vulnerabilities can be shared openly within the community without them falling into the `wrong' hands}. 

We have performed two preliminary interview studies to assess how software engineers and end-user organizations would use trust data in their work. These interview studies showed that while trust facts may be irrefutable, the perception of trust is a moving target for different actors. One actor may be satisfied with using a bleeding-edge version of a software package that updates often, while another wants an old version of a patched but reliable package. We are currently exploring if \textbf{different categorizations of trust ratings are necessary}, such as ``project evolution speed'', ``vulnerability fix time'', and ``team liveliness''. 

While we respect tools that centrally collect trust data, we believe that trust data should be part of a zero-trust community. We observe that it is \textbf{impossible to observe correct trust data as one central party objectively} and consensus mechanisms are necessary to observe that a software package version is trustworthy objectively.

The data collection process itself is challenging. We should be cognizant that a set of impact factors under security in Figure \ref{fig:factors}, for instance, will be challenging to operationalize. Once we have identified the values (i.e., trust facts) necessary to collect, we will have to tackle the challenge of the availability of these data. Data needed to judge the software trust can come from usage data, source code, build systems, test systems, etc. It will thereby be challenging to \textbf{judge the trust of closed software}. Furthermore, it is hard, if not impossible, to \textbf{get data from closed reputation databases}, such as app store ratings. Finally, the \textbf{operationalization of the impact factors} is a major challenge in this work. We plan to perform a large-scale survey study with software engineers as an approach to this problem to identify a minimal collectible set of trust facts useful for software engineers.

\subsection{Future Work}
Our near-future works consist roughly of two parts.

First, we plan to \textbf{launch a software engineer survey based on the outcomes of this SLR}. The developer survey intends to determine a prioritized and categorized list of trust criteria. Furthermore, we want to use the survey to gauge and create awareness amongst software engineers about the role of trust in their software engineering processes.

Secondly, using the list of prioritized trust factors, \textbf{we develop a tool for automatically gathering trust scores for different trust impact categories}~(\cite{hou2021trustseco}). Software engineers must evaluate the tool to decide whether it effectively supports them in downloading and depending on a particular software package. Finally, we aim to store these trust scores in a distributed ledger, to make trust common for all software engineers. The community that uses the ledger can contribute openly observable facts to the ledger, to increase or decrease trust scores for software packages.

\section{Conclusion}
\label{conclusion}

Although prior work has developed a number of models for quantifying software trust or other related attributes, only a few have been widely used in real-world projects. We discovered that the majority of them are oriented around software quality measurement, ignoring the impact of related actors in SECO on trust, and omitting the role of attacks and vulnerabilities in measuring trust.

In this paper, we examined the role of trust in SECO by starting with the selection and measurement of software products. Our goal is to devise a strategy for resolving the trust erosion issue raised in the introduction, specifically the harm caused to SECO's health by vulnerabilities and attacks. This study adopted a systematic literature review approach to review trust in SECO from a selected 112 articles. On this basis, we discussed the definition, types, and sources of software trust and proposed a definition for software trust in the context of SECO. Additionally, we classified the impact factors on software trust from the standpoints of software and packages, software package managers, software producers, and software end-users. We analyzed the existing literature for relevant impact factors and counted their frequencies. Software quality, SPO, code \& structure, documentation, and structural assurance are the top five impact factors. Furthermore, based on these studies, we compiled a comprehensive table of software trust impact factors and identified as many practical metrics as possible for each factor to be measured. 

Although the majority of the trust factors we collected are based on empirical research, and only a few models or metrics are widely used in real-world projects, they reflect, to some extent, the concerns expressed by software engineers or end-users during the software selection or evaluation process.

According to the literature, software trust serves as the foundation for cooperation between hubs and stakeholders in SECO, while software trust is critical for cooperation between actors, current research on software trust focuses exclusively on software and components, particularly software quality. Limited measures of trust assessment cannot be the final word to success, it should be expanded to include other entities to analyze trust in the software ecosystem in a multi-level, multi-perspective manner. Next, we will conduct a survey of software engineers to ascertain their perceptions of software trust in the future, paving the way for the design of a community-managed infrastructure that serves as a trust layer for SECO.

\begin{acknowledgements}
We thank the SecureSECO team for their constructive feedback and help in the creation of this article. In particular, we thank Donny Groeneveld, Venja Beck, and Floris Jansen for their diligence in performing their part of the literature review, i.e., filtering, double-checking, and processing the articles. Furthermore, we thank them for the lengthy and productive discussions that led to an inter-rater agreement between the SLR team members. Finally, we thank Hidde Reeskamp and Joost Gadellaa for their excellent comments on the early versions of this article.
\end{acknowledgements}



\printbibliography[keyword=primary,heading=subbibliography,%
title={\textbf{Systematic Literature Review Articles}}]
\printbibliography[keyword=secondary,heading=subbibliography,%
title={\textbf{Other Sources}}]

\end{document}